\documentclass[aps,pra,reprint,superscriptaddress,nopacs,nofootinbib,longbibliography]{revtex4-1}

\usepackage{amsmath,amssymb,latexsym,booktabs}
\usepackage{graphicx}
\usepackage{hyperref}
\usepackage{mathrsfs}
\usepackage{amscd}
\usepackage{color}
\usepackage{comment}
\DeclareSymbolFont{symbols}{OMS}{cmsy}{m}{n}
\DeclareSymbolFont{largesymbols}{OMX}{cmex}{m}{n}
\renewcommand{\bm}[1]{\boldsymbol #1}

\begin{document}

\title{
Strong eigenstate thermalization in Singer-cycle Clifford circuits
}

\author{Naoto Tsuji}
\affiliation{Department of Physics, University of Tokyo, Hongo, Tokyo 113-0033, Japan}
\affiliation{RIKEN Center for Emergent Matter Science (CEMS), Wako 351-0198, Japan}
\affiliation{Trans-scale Quantum Science Institute, University of Tokyo, Hongo, Tokyo 113-0033, Japan}

\begin{abstract}
Eigenstate thermalization hypothesis (ETH) provides a central framework for understanding thermalization in isolated quantum systems. 
Recently, the infinite-temperature ETH has been established for almost all Floquet eigenstates in linear feedback shift register quantum circuits.
Here we introduce brickwork Clifford circuits with $L$ qubits where the induced transformation of Pauli operators is represented by an irreducible symplectic matrix over the finite field $\mathbb F_2$ of the maximal order $2^L+1$, known as a Singer cycle.
For these Singer-cycle Clifford circuits, we prove a strong version of ETH:
{\it all} Floquet eigenstates are thermal at infinite temperature without exceptions. 
We derive uniform bounds on both diagonal 
deviations from thermal expectation values 
and off-diagonal matrix elements of
arbitrary observables supported on a fixed number of sites.
The results indicate that long-time averages of observables approach the infinite-temperature values in the thermodynamic limit for arbitrary initial states. 
We also show that the temporal fluctuation of expectation values of observables is exponentially suppressed in the thermodynamic limit, implying that each observable remains close to the infinite-temperature value for almost all times.
\end{abstract}


\date{\today}

\maketitle

\section{Introduction}
\label{sec: introduction}

Understanding the mechanism of thermalization in isolated quantum systems lies at the fundamental core of thermodynamics and statistical mechanics in quantum systems \cite{Gogolin2016, DAlessio2016}.
While we empirically know that expectation values of observables approach a thermal value if a quantum system is left as it is for a long time, it is not obvious how it arises from unitary dynamics where the system's density matrix does not approach a thermal one in the operator level. Among various approaches, eigenstate thermalization hypothesis (ETH) \cite{Deutsch1991, Srednicki1994, RigolDunjkoOlshanii2008, Deutsch2018, Mori2018} offers a compelling perspective on this issue:
thermal properties are already encoded in individual eigenstates of the Hamiltonian.
If ETH holds, one can immediately show that 
the time average of expectation values of observables approach thermal values at the corresponding energy.

\begin{figure}[ht]
\includegraphics[width=8.5cm]{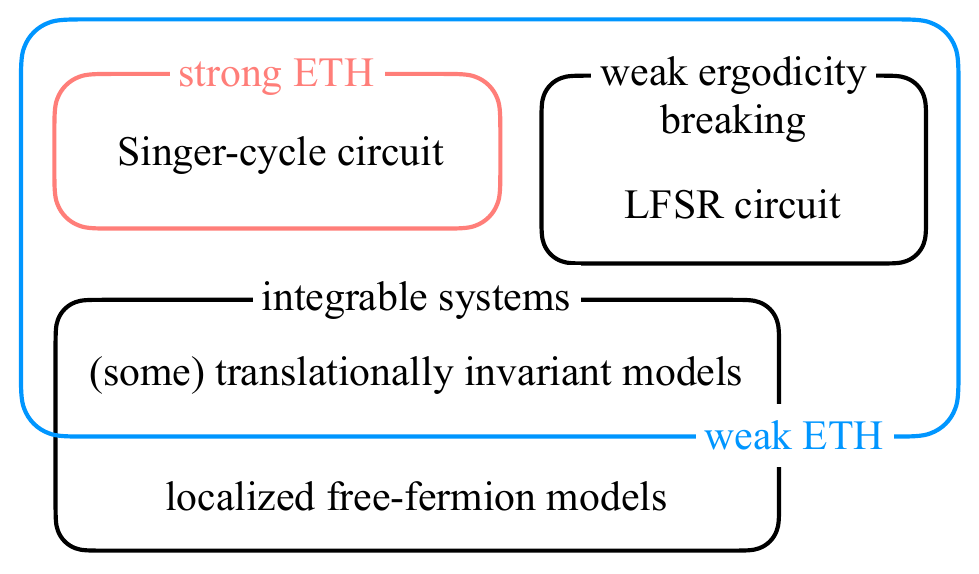}
\caption{Examples of quantum many-body systems classified according to the strong and weak ETH. 
The strong ETH requires all eigenstates in the relevant energy window and symmetry sector to be thermal, 
while the weak ETH allows a fraction of nonthermal eigenstates that vanishes in the thermodynamic limit. 
Here LFSR circuit refers to the one studied in Ref.~\cite{IppolitiLong2026}. 
Integrable systems can satisfy or violate the weak ETH: representative translation-invariant models can satisfy it, whereas spatially localized free-fermion models violate it \cite{Lydzba2024}.}
\label{fig:eth_classification}
\end{figure}

While there have been various numerical studies that support ETH \cite{RigolDunjkoOlshanii2008, DAlessio2016}, it has been challenging to show ETH rigorously, even for a specific quantum many-body model. There is a weaker version of ETH, stating that almost all eigenstates of the Hamiltonian are thermal (i.e., their expectation values of observables are close to the thermal ones) \cite{Biroli2010, Iyoda2017, Lydzba2024}. This requires only that the fraction of thermal eigenstates approach unity in the thermodynamic limit.
The weak ETH can be established under broad conditions, including some integrable systems, and therefore does not distinguish thermalizing systems from non-thermalizing ones (Fig.~\ref{fig:eth_classification}). 
This motivates a sharper estimate on the number of thermal eigenstates, or equivalently, of exceptional nonthermal eigenstates.

A strong version of ETH excludes such exceptions \cite{Kim2014, Yoshizawa2018}:
all eigenstates within the relevant energy window and symmetry sector are thermal,
and hence long-time averages of observables approach the thermal values
for initial states confined to those regimes.
More precisely, the strong ETH states that the expectation values of an observable $O$ for all energy eigenstates
in each symmetry sector within the energy range of interest approach their corresponding thermal values
uniformly:
\begin{align}
\left|
\langle\psi_k|O|\psi_k\rangle-O_{\rm th}(E_k)
\right|
&\le \varepsilon(L),
\end{align}
where $E_k$ is the eigenenergy of $|\psi_k\rangle$ and
$\varepsilon(L)$ is independent of the eigenstate label $k$
and vanishes as the system size $L$ tends to infinity.
A more detailed ETH ansatz describes both diagonal and
off-diagonal matrix elements, including the scaling and
statistical properties of the fluctuations \cite{DAlessio2016, Srednicki1999}.
For periodically driven (Floquet) systems \cite{Oka2019, Tsuji2024}, thermalization is expected to lead to an infinite-temperature state.
The corresponding Floquet ETH states that expectation values of observables
in Floquet eigenstates approach the infinite-temperature
expectation values \cite{Lazarides2014, DAlessio2014}.

There have been various theoretical studies towards establishing thermalization and ETH in isolated quantum systems. Among them, in Ref.~\cite{Chiba2024} certain nonintegrable spin chains were shown to possess exact eigenstates that have infinite-temperature expectation values for spatially local observables.
This approach was extended to a broader class of stabilizer states \cite{Hokkyo2026}.
Rigorous results on macroscopic equilibration have been obtained for coarse-grained particle densities in free-fermion systems \cite{ShiraishiTasaki2024, Tasaki2025, HaraKoike2026}.

\begin{figure*}[t]
\includegraphics[width=18cm]{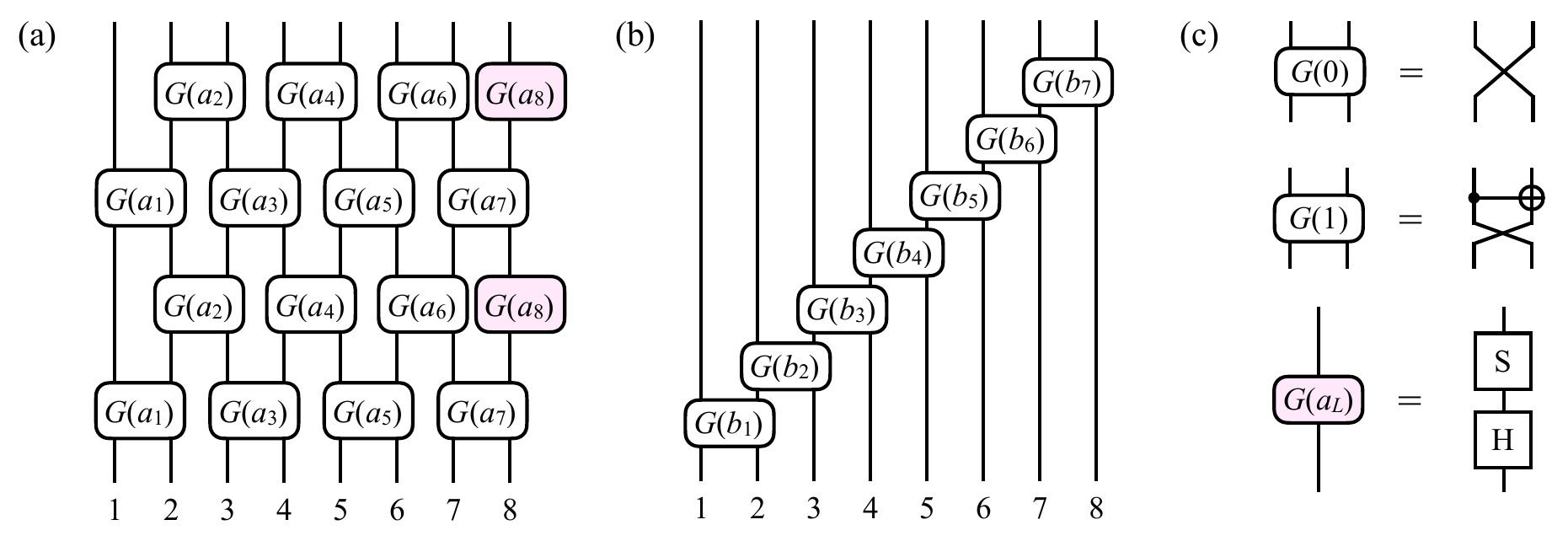}
\caption{
Circuit diagrams of (a) a Singer-cycle Clifford circuit and (b) an LFSR circuit illustrated for $L=8$.
(c) Representation of the bulk gates $G(a_i)$ and $G(b_i)$: 
each gate consists of a SWAP gate when the binary coefficient is $0$, 
and a combination of SWAP and CNOT gates when it is $1$.
The endpoint gate $G(a_L)$ marked by pink (where we always choose $a_L=1$) in (a) consists of phase (S) and Hadamard (H) gates.
The coefficient sets $\{a_i\}$ and $\{b_i\}$ are chosen according to different criteria.}
\label{fig:circuit}
\end{figure*}

Another recent work, which is closely related to the present paper, has shown that for a family of quantum circuits implementing linear feedback shift registers (LFSRs) all Floquet eigenstates except for those in a two-dimensional eigenspace obey the infinite-temperature ETH \cite{IppolitiLong2026}.
As we will see, these circuits have no local conserved quantity, so the exceptional nonthermal eigenstates can be viewed as quantum many-body scar states \cite{Bernien2017, Shiraishi2017, Turner2018, Serbyn2021, Moudgalya2022}.
In scarred systems, although the number of nonthermal eigenstates is negligible,
an initial state may have substantial overlap with nonthermal eigenstates,
thereby preventing thermalization (weak ergodicity breaking, Fig.~\ref{fig:eth_classification}) \cite{Turner2018, Serbyn2021}.
Motivated by these developments, we address the following question: Can one construct an explicit quantum many-body model for which the strong ETH can be rigorously established, without any exceptional nonthermal eigenstates, and hence observables reach thermal values irrespective of the choice of initial states? 

In this paper, we consider brickwork Clifford circuits composed of CNOT and SWAP gates in the bulk and Hadamard and phase gates at the right endpoint (see Fig.~\ref{fig:circuit}(a), with each gate represented in (c)). 
While the LFSR circuits (Fig.~\ref{fig:circuit}(b)) permute computational-basis states,
the endpoint gates in our circuit generate superpositions and relative phases among them, allowing for richer state dynamics.
On the operator side, the endpoint gates mix $X$ and $Z$ Pauli operators (distinction between shift-register and brickwork arrangements is not important \cite{Clifford_ordering}).
We choose the gate coefficients so that the induced transformation of Pauli operators is represented by an irreducible symplectic matrix over the finite field $\mathbb F_2$ with the maximal order $2^L+1$ ($L$ is the number of qubits), known as a Singer cycle \cite{Singer1938, Huppert1970, Bereczky2000}.
Consequently, every (non-identity) Pauli operator has an orbit
of length $2^L+1$, up to phase factors.

For these Singer-cycle Clifford circuits, we rigorously establish the strong ETH without any exceptional nonthermal eigenstates (Fig.~\ref{fig:eth_classification}).
Despite the resemblance to the LFSR circuits, the proof of the strong ETH for the present circuits is not a straightforward extension, since the circuit dynamics can no longer be described as a permutation of computational-basis states. Instead of explicitly writing down the eigenstates on the computational basis, we keep track of the evolution of Pauli operators and describe squared matrix elements of observables in terms of finite fields.

Using the Weil bound on mixed character sums over finite fields \cite{Weil1948, Castro2000, Cochrane2006}, we derive uniform bounds of order $D^{-1/2}$ \cite{bound_scaling} on both diagonal deviations from infinite-temperature expectation values and
off-diagonal matrix elements of observables $O$ supported on a fixed number of sites,
\begin{align}
\left|\langle\psi_k|O|\psi_{k'}\rangle-\frac{\operatorname{Tr}O}{D}\delta_{k,k'}
\right|
&\le
c D^{-1/2},
\label{eq:infinite_T_ETH}
\end{align}
where $c$ is a constant that does not depend on the system size and the eigenstate labels $k, k'$, and $D=2^L$ is the Hilbert-space dimension.
We further show that, for any initial state, the temporal fluctuations of these expectation values from the thermal values vanish in the thermodynamic limit. Thus, each observable remains close to its infinite-temperature value for almost all times, regardless of the choice of initial states.
As a byproduct of the temporal fluctuation analysis, we obtain an alternative (and simpler) proof of the strong ETH but with a weaker bound of order $D^{-1/4}$.

The paper is organized as follows. 
In Sec.~\ref{sec:LFSR_circuit}, we review the LFSR circuits studied in Ref.~\cite{IppolitiLong2026}. 
In Sec.~\ref{sec:Singer_cycle}, we introduce Singer-cycle Clifford circuits and describe 
their properties, including the characteristic polynomial, gate coefficients, eigenvalues and eigenstates of the Floquet operator, and phase convention in terms of finite fields.
In Sec.~\ref{sec:strong_ETH}, we establish the strong ETH by bounding matrix elements of Pauli operators using the Weil bound on finite-field character sums. 
In Sec.~\ref{sec:temporal_fluctuation}, we derive bounds on temporal fluctuations of expectation values of observables. 
Finally, Sec.~\ref{sec:summary} summarizes the paper.

\section{LFSR circuits}
\label{sec:LFSR_circuit}

In this section, we briefly review the LFSR circuit developed by Ippoliti and Long \cite{IppolitiLong2026}, 
for which ETH holds for almost all eigenstates (except for two degenerate eigenstates) under suitable conditions on the circuit coefficients.

In the LFSR circuit (Fig.~\ref{fig:circuit}(b)),
the unitary operator generating one period of time evolution (Floquet operator) takes the form of
\begin{align}
U
&=
G_{L-1}(b_{L-1}) G_{L-2}(b_{L-2}) \cdots G_1(b_1),
\label{eq:LFSR_U}
\end{align}
where $b_i=0, 1\in\mathbb F_2$ are binary coefficients ($\mathbb F_2=\{0,1\}$ denotes the finite field of characteristic 2) and
\begin{align}
G_i(b_i)
&=
({\rm CNOT}_{i,i+1})^{b_i} {\rm SWAP}_{i,i+1} 
\notag
\\
&=
{\rm SWAP}_{i,i+1} ({\rm CNOT}_{i+1,i})^{b_i} 
\quad
(i=1, 2, \dots, L-1)
\label{eq:bulk_gate}
\end{align}
are gate operators.
$\mathrm{CNOT}_{i,j}$ denotes a controlled-NOT
gate with control qubit $i$ and target qubit $j$,
and ${\rm SWAP}_{i,j}$ represents a swap operator that exchanges 
states at sites $i$ and $j$.
The action of each gate on the two-site computational
basis is given by
$G_i(b_i)|z_i,z_{i+1}\rangle=|z_{i+1},z_i+b_i z_{i+1}\rangle$,
where addition and multiplication is performed in $\mathbb F_2$.

We can write the total computational basis in terms of a bit string
as $|\bm z\rangle$ with $\bm z=(z_1,\dots,z_L)^T\in\mathbb F_2^L$.
In the LFSR circuit,
a state $|\bm z\rangle$ evolves after one period as $U|\bm z\rangle=|A\bm z\rangle$,
where $A$ is an $\mathbb F_2$-valued matrix defined by
\begin{align}
A
&=
\begin{pmatrix}
0 & 1 & 0 & \cdots & 0 \\
0 & 0 & 1 & \cdots & 0 \\
\vdots & \vdots & \ddots & \ddots & \vdots \\
0 & 0 & \cdots & 0 & 1 \\
1 & b_1 & b_2 & \cdots & b_{L-1}
\end{pmatrix}
\in \mathrm{GL}(L,\mathbb F_2).
\end{align}
Thus, the bits are shifted as $z'_i=z_{i+1}$ for $i<L$ and 
the end bit changes as
$z'_L=z_1+\sum_{i=1}^{L-1}b_i z_{i+1}$.
One can see that 
$U$ permutes the computational-basis states without introducing phases or superpositions,
which allows one to construct the eigenstates of the Floquet operator 
directly from the orbits of bit strings.

The characteristic polynomial of $A$ is
\begin{align}
g_L(t)
&=
\det(tI-A)
=
t^L+\sum_{j=1}^{L-1} b_j t^j+1.
\end{align}
To derive ETH, the gate coefficients $b_j$ are chosen in such a way that $g_L(t)$ becomes
a primitive polynomial over $\mathbb F_2$.
That is, $g_L(t)$ is irreducible, and every root of $g_L(t)$ has multiplicative order $2^L-1$.
These conditions ensure that an orbit starting from
a nonzero bit string, $\bm z_0\to A\bm z_0\to A^2\bm z_0\to \cdots$,
forms a periodic cycle with a period,
\begin{align}
N
&=
D-1,
\end{align}
where $D=2^L$ is the dimension of the Hilbert space.
In particular, $A^N=I$ and hence $U^N=I$, implying that the
eigenvalues of $U$ are $N$th roots of unity,
\begin{align}
\omega^k
&=
e^{2\pi ik/N}
\quad
(k=0,1,\dots,N-1).
\end{align}
By choosing an arbitrary nonzero bit string $\bm z_0$,
one can construct $N$ eigenstates of $U$ as
\begin{align}
|\psi_k\rangle
&=
\frac{1}{\sqrt{N}} \sum_{j=0}^{N-1} \omega^{jk} |A^j\bm z_0\rangle
\quad
(k=0,1,\dots,N-1)
\end{align}
with eigenvalues $\omega^{-k}$. 
They are orthogonal to each other ($\langle \psi_k|\psi_{k'}\rangle=\delta_{k,k'}$).

The remaining eigenstate is $|\bm 0\rangle=|0\rangle^{\otimes L}$, which is invariant under $U$ ($U|\bm 0\rangle=|\bm 0\rangle$).
Thus, the eigenvalue $1$ is twofold degenerate,
with the eigenspace spanned by $|\psi_0\rangle$ and
$|\bm 0\rangle$, while all other eigenvalues are nondegenerate.
The degenerate eigenspace is also spanned by $|0\rangle^{\otimes L}$ and $|+\rangle^{\otimes L}$,
since $|+\rangle^{\otimes L}=\frac{1}{\sqrt{D}}|\bm 0\rangle+\sqrt{\frac{D-1}{D}}|\psi_0\rangle$.
The states $|0\rangle^{\otimes L}$ and $|+\rangle^{\otimes L}$ are nonthermal eigenstates,
since they have nonzero expectation values for some Pauli operators (e.g., ${}^{\otimes L}\langle 0|Z_j|0\rangle^{\otimes L}=1$, ${}^{\otimes L}\langle +|X_j|+\rangle^{\otimes L}=1$).
Recall that all (non-identity) Pauli operators have vanishing expectation values at the infinite-temperature thermal state.

Since $g_L(t)$ is irreducible over $\mathbb F_2$, one can define a finite field,
\begin{align}
\mathbb F_2[t]/g_L(t)
\simeq
\mathbb F_D.
\end{align}
Let $\alpha$ be a root of $g_L(t)$.
Since $g_L(t)$ is primitive, $\alpha$ becomes a generator of the multiplicative group $\mathbb F_D^\times
=\{1, \alpha, \alpha^2, \dots, \alpha^{N-1}\}$.
For any nonzero $\bm z_0\in\mathbb F_2^L$,
the vectors $\bm z_0,A\bm z_0,\dots,A^{L-1}\bm z_0$
form a basis of $\mathbb F_2^L$.
One can identify the computational-basis label space $\mathbb F_2^L$
with the finite field $\mathbb F_D$ through the mapping
\begin{align}
\bm z
&=
\sum_{j=0}^{L-1} c_j A^j \bm z_0
\mapsto
z
=
\sum_{j=0}^{L-1} c_j \alpha^j.
\label{eq:z_vec_z}
\end{align}
Under this identification, the action of $A$ becomes multiplication by $\alpha$.
If we write $|z\rangle$ for the computational-basis state $|\bm z\rangle$, we obtain
$U|z\rangle=|\alpha z\rangle$. Note that this mapping is $\mathbb F_2$-linear,
i.e., $c_1\bm z_1+c_2\bm z_2$ is mapped to $c_1z_1+c_2z_2$
($c_1, c_2\in\mathbb F_2$).

Let us define the multiplicative character of $\mathbb F_D^\times$ as
\begin{align}
\chi_k(\alpha^n)
&=
\omega^{kn}
\quad
(k=0,1,\dots,N-1),
\end{align}
which satisfies $\chi_k(g)\chi_k(g')=\chi_k(gg')$,
$\chi_k(g)\chi_{k'}(g)=\chi_{k+k'}(g)$,
and $\overline{\chi_k(g)}=\chi_{-k}(g)$ for $g, g'\in \mathbb F_D^\times$
(the overline denotes complex conjugation).
With this notation, one can write the eigenstates of $U$ as
\begin{align}
|\psi_k\rangle
&=
\frac{1}{\sqrt{N}}
\sum_{z\in \mathbb F_D^\times} \chi_k(z)|z\rangle.
\label{eq:psi_k_expansion}
\end{align}

Matrix elements of Pauli operators in the eigenstate basis can be expressed
in terms of the characters of the finite field $\mathbb F_D$.
For $\bm u,\bm v\in\mathbb F_2^L$, we define
\begin{align}
X(\bm u)
&=
\bigotimes_{i=1}^L X_i^{u_i},
\quad
Z(\bm v)
=
\bigotimes_{i=1}^L Z_i^{v_i}.
\end{align}
where $X_i$ and $Z_i$ are Pauli operators acting on site $i$.
Arbitrary operator can be expressed as a linear combination of $X(\bm u)Z(\bm v)$.
The actions of $X(\bm u)$ and $Z(\bm v)$ on the computational basis are given by
\begin{align}
X(\bm u)|\bm z\rangle
&=
|\bm z+\bm u\rangle,
\\
Z(\bm v)|\bm z\rangle
&=
(-1)^{\bm v\cdot\bm z}|\bm z\rangle,
\end{align}
respectively.

Let us define the additive character of $\mathbb F_D$ by
\begin{align}
\varphi_v(z)
&=
(-1)^{\bm v\cdot\bm z}
\quad
(v, z\in \mathbb F_D),
\end{align}
where $v$ and $z$ are the field labels corresponding
to $\bm v$ and $\bm z$ ($\in\mathbb F_2^L$) through the mapping \eqref{eq:z_vec_z}, respectively.
It satisfies $\varphi_v(z+z')=\varphi_v(z)\varphi_v(z')$.
Using the characters $\chi$ and $\varphi_v$,
one can write the matrix elements of Pauli operators in the computational basis as
\begin{align}
\langle z|X(u)Z(v)|z'\rangle
&=
\varphi_v(z')\delta_{z,z'+u}.
\end{align}
Substituting the expansions of the eigenstates \eqref{eq:psi_k_expansion},
one can obtain the matrix elements of Pauli operators in the eigenstate basis as
\begin{align}
\langle \psi_k|X(u)Z(v)|\psi_{k'}\rangle
&=
\frac{1}{N}{\sum_{z\in \mathbb F_D}}'
\chi_{-k}(z+u)\chi_{k'}(z)\varphi_v(z),
\end{align}
where the sum $\sum'$ is taken over $\mathbb F_D$ except for $z=0$ and $u$.
If one defines $\chi(z):=\chi_1(z)$ and uses the relation $\chi_k(g)=\chi(g^k)$,
one arrives at the expression
\begin{align}
\langle \psi_k|X(u)Z(v)|\psi_{k'}\rangle
&=
\frac{1}{N}{\sum_{z\in \mathbb F_D}}'
\chi(z^{k'}(z+u)^{-k})\varphi_v(z).
\label{eq:Pauli_eval}
\end{align}

Now we assume that $k, k'\neq 0$ and $u, v\neq 0$. In this case, the characters in Eq.~\eqref{eq:Pauli_eval} become nontrivial
(i.e., they are not identity characters),
and Eq.~\eqref{eq:Pauli_eval} takes the form of the sum of mixed characters (i.e., a product of 
multiplicative and additive characters), which is known to have a suitable bound (Weil bound) \cite{Weil1948, Castro2000, Cochrane2006},
\begin{align}
\left|
{\sum_{z\in \mathbb F_D}}'
\chi(z^{k'}(z+u)^{-k})\varphi_v(z)
\right|
&\le
2\sqrt{D}
\quad
(k,k'\neq 0).
\end{align}
As a result, one can bound the matrix elements of Pauli operators in the eigenstate basis 
(except for two degenerate eigenstates):
\begin{align}
|\langle \psi_k|X(u)Z(v)|\psi_{k'}\rangle|
&\le
\frac{2\sqrt{D}}{D-1}
\quad
(k,k'\neq 0).
\label{eq:Pauli_eval2}
\end{align}
Similar bounds can be obtained in the cases of $(u=0, v\neq 0)$ and $(u\neq 0, v=0)$.
Combining these results leads to the statement of the infinite-temperature ETH \eqref{eq:infinite_T_ETH}
for almost all eigenstates
for the LFSR circuit. For the derivation from Eq.~\eqref{eq:Pauli_eval2} to Eq.~\eqref{eq:infinite_T_ETH}, 
we refer to Ref.~\cite{IppolitiLong2026}
(see also Sec.~\ref{sec:mixed_sum}) .

Let us note that in the LFSR circuit there are no nontrivial local conserved quantities.
To see this, it is convenient to use the fact that Pauli operators are transformed by $U$ as $UX(\bm u)Z(\bm v)U^\dagger=X(A\bm u)Z((A^{-1})^T\bm v)$.
Suppose that there exists a local conserved quantity $Q$.
If one expands $Q$ in the Pauli-operator basis, 
the relation $UQU^\dagger=Q$ implies that its coefficients are constant across each orbit of Pauli operators. Consequently, if a given Pauli operator appears in the expansion of $Q$, then the entire orbit to which it belongs appears with the same coefficient.
In the LFSR circuit, the length of each orbit of non-identity Pauli operators is $D-1$. On the other hand, if each local Pauli operator acts on at most $\ell$ adjacent sites, the number of terms appearing in $Q$ is at most $L(4^\ell-1)$.
Since $L(4^\ell-1)<D-1$ for sufficiently large systems, we conclude that there is no such kind of $Q$, implying the absence of local conserved quantities.
In particular, the exceptional eigenspace spanned by
$|0\rangle^{\otimes L}$ and $|+\rangle^{\otimes L}$ that do not obey ETH is not
distinguished from the thermal eigenstates by any local conserved quantity. 

There are three features in the LFSR circuit that are relevant to our construction in the next section. 
First, the CNOT and SWAP gates just permute the computational-basis states without introducing phases
or superpositions, so the eigenstates can be constructed directly from the orbits
of computational-basis states.
Second, one can choose the characteristic polynomial $g_L(t)$ to be a primitive polynomial, which allows one to realize
a single orbit of maximal length $2^L-1$.
Third, the ETH can be shown for almost all eigenstates except for two degenerate eigenstates.
In the present work, we will construct a Floquet circuit model that allows us to show the strong ETH \eqref{eq:infinite_T_ETH}, i.e.,
all the eigenstates obey the ETH without exceptions.
To this end, we have to find a different way to evaluate matrix elements of observables without relying on the computational-basis representation,
for which we need a different type of characteristic polynomials.

\section{Singer-cycle Clifford circuits}
\label{sec:Singer_cycle}

In this section, we construct a family of brickwork Floquet Clifford circuits (Fig.~\ref{fig:circuit}(a)) for which every eigenstate satisfies ETH
at infinite temperature \eqref{eq:infinite_T_ETH}. The key difference from the LFSR circuit is an endpoint gate that combines Hadamard and phase gates. 
The Hadamard gate mixes computational-basis states, so the bit-string orbit construction of the previous section no longer applies. 
Since the circuit is composed of Clifford gates,
the Floquet operator maps each Pauli operator to another Pauli operator up to a phase. 
Hence we keep track of Pauli operator strings instead of bit strings of computational-basis states.
We first derive the action of the Floquet operator on Pauli operators, and then choose the gate coefficients 
so that the transformation of Pauli operator strings is represented by an irreducible symplectic matrix
of order $2^L+1$ ($L$ is the number of qubits), which is known as a Singer cycle \cite{Singer1938, Huppert1970, Bereczky2000}.
Thus, every (non-identity) Pauli operator has an orbit of length $2^L+1$,
which underlies our analysis of matrix elements of observables
for the proof of the strong ETH.

\subsection{Gate operation on Pauli operators}

Let us consider a chain of $L$ qubits with Hilbert space dimension $D=2^L$.
For the brickwork circuit shown in Fig.~\ref{fig:circuit}(a), the Floquet operator takes the form
\begin{align}
U
&=
\prod_{\substack{1\leq i\leq L\\ i {\rm : even}}}
G_i(a_i)
\prod_{\substack{1\leq i\leq L\\ i {\rm : odd}}}
G_i(a_i),
\label{eq:brickwork_matrix}
\end{align}
where $a_i\in\{0, 1\}=\mathbb F_2$ are binary coefficients,
\begin{align}
G_i(a_i)
&=
({\rm CNOT}_{i,i+1})^{a_i} {\rm SWAP}_{i,i+1} 
\notag
\\
&=
{\rm SWAP}_{i,i+1} ({\rm CNOT}_{i+1,i})^{a_i}
\quad
(i=1, 2, \dots, L-1)
\end{align}
corresponds to a bulk gate action,
and 
\begin{align}
G_L(a_L)
&=
S^{a_L}H
\label{eq:endpoint_gate}
\end{align}
denotes an endpoint gate acting on site $L$.
Here $S=\operatorname{diag}(1,i)$ is the phase gate,
and $H=\frac{1}{\sqrt{2}}(X+Z)$ is the Hadamard gate.
We employ the same definition for the bulk gates as in Eq.~\eqref{eq:bulk_gate}. 

To describe its action on operators, we introduce Pauli strings in the form of
\begin{align}
P(\bm v)
&=
X^{\bm x}Z^{\bm z}
=
\bigotimes_{i=1}^{L}X_i^{x_i}Z_i^{z_i},
\label{eq:pauli_definition}
\end{align}
where
\begin{align}
\bm v
&=
\begin{pmatrix}
\bm x\\
\bm z
\end{pmatrix}
\in\mathbb F_2^{2L}
\end{align}
is the label of the Pauli strings.
Here $\bm x,\bm z\in\mathbb F_2^L$, and all operations on these
labels are performed over $\mathbb F_2$.
Any operator can be expressed as a linear combination of Pauli strings.
For example, local operators acting on a single qubit can be written as a linear combination of
$I=X^0 Z^0$, $X$, $Y=iXZ$, and $Z$.
We use the convention of Pauli strings [Eq.~(\ref{eq:pauli_definition})]
without an additional phase factor. Consequently, $P(\bm v)$ is unitary
but need not be Hermitian.

Since the Floquet operator $U$ is a Clifford unitary,
the unitary evolution by $U$ maps each Pauli string to another
Pauli string up to a phase,
\begin{align}
UP(\bm v)U^\dagger
&=
\eta(\bm v)P(M\bm v),
\label{eq:clifford_action}
\end{align}
where the binary matrix $M\in {\rm GL}(2L, \mathbb F_2)$ describes the evolution of the Pauli string labels
and
\begin{align}
\eta(\bm v)\in\{1,-1,i,-i\}
\end{align}
is a phase factor
in the convention of Eq.~\eqref{eq:pauli_definition}.

For a bulk gate $G_i(a_i)$ acting on sites $i$ and $i+1$, the Pauli string labels are transformed as
\begin{align}
x_i'
&=
x_{i+1},
\\
x_{i+1}'
&=
x_i+a_i x_{i+1},
\notag\\
z_i'
&=
z_{i+1}+a_iz_i,
\\
z_{i+1}'
&=
z_i.
\label{eq:bulk_label_update}
\end{align}
Thus, the bulk gates transform the $X$ and $Z$ labels separately.
The corresponding matrix acting on the labels $(x_i,x_{i+1},z_i,z_{i+1})$ is explicitly written as
\begin{align}
R_i(a_i)
&=
\begin{pmatrix}
R^{(1)}(a_i) & 0 \\
0 & R^{(2)}(a_i)
\end{pmatrix}
\quad (i=1,\ldots,L-1),
\label{eq:bulk_label_matrix}
\end{align}
where we define $2\times 2$ matrices over $\mathbb F_2$,
\begin{align}
R^{(1)}(a)
&=
\begin{pmatrix}
0 & 1 \\
1 & a
\end{pmatrix}
\quad
(a\in \mathbb F_2),
\\
R^{(2)}(a)
&=
\begin{pmatrix}
a & 1 \\
1 & 0
\end{pmatrix}
\quad
(a\in \mathbb F_2).
\label{eq:bulk_blocks}
\end{align}
Each $R_i$ is understood as a $2L\times2L$ matrix acting
as the identity on all remaining components.
At the right endpoint, we apply $G_L$ (Eq.~(\ref{eq:endpoint_gate})), whose action on the label $(x_L,z_L)$ is given by
\begin{align}
R_L(a_L)
&=
\begin{pmatrix}
0 & 1 \\
1 & a_L
\end{pmatrix}
=
R^{(1)}(a_L).
\label{eq:boundary_label_matrix}
\end{align}
The choices $a_L=0$ and $a_L=1$ correspond to $H$ and $SH$,
respectively.
Unlike the bulk gates, this endpoint gate mixes the $X$ and $Z$
labels.
Using these $R_i$'s, we can write the label transformation matrix $M$ in Eq.~(\ref{eq:clifford_action}) as
\begin{align}
M
&=
\prod_{\substack{1\leq i\leq L\\ i{\rm : even}}}
R_i(a_i)
\prod_{\substack{1\leq i\leq L\\ i{\rm : odd}}}
R_i(a_i).
\label{eq:matrix_M}
\end{align}

In our phase convention, the bulk gates acting on Pauli strings do not induce an additional phase factor,
\begin{align}
G_i(a_i) P(\bm v) G_i(a_i)^\dagger
&=
P(R_i(a_i)\bm v)
\quad
(i=1,2,\dots,L-1).
\end{align}
A nontrivial phase arises only from the endpoint gate action.
Since
\begin{align}
S^{a}X^x Z^z S^{a\dagger}
&=
i^{ax} X^x Z^{z+ax},
\notag
\\
HX^xZ^z H^\dagger
&=
(-1)^{xz} X^z Z^x,
\end{align}
a Pauli string transforms as
\begin{align}
G_L(a_L)P(\bm v)G_L(a_L)^\dagger
&=
\eta_L(\bm v)\,P(R_L(a_L)\bm v)
\end{align}
with a phase factor
\begin{align}
\eta_L(\bm v)
&=
i^{a_L z_L}(-1)^{x_Lz_L}.
\end{align}
Here we remark that the labels of Pauli strings (such as $x_L$, $z_L$, and $a_L z_L$) are evaluated in $\mathbb F_2$, 
while the exponent of $i$ should be understood as the representatives
$a_L, z_L\in\{0,1\}$ (for example, $1\equiv 3$ (mod 2) but $i^1 \neq i^3$).
One should also note that in the brickwork circuit $\eta_L$ is evaluated
at the label immediately before the endpoint gate acts,
including the transformations induced by the preceding
bulk gates.

The matrix $M$ (\ref{eq:matrix_M}) preserves the commutation relations of Pauli strings.
For $\bm v=(\bm x,\bm z)^T$ and
$\bm v'=(\bm x',\bm z')^T$,
Pauli strings are commuted as
\begin{align}
P(\bm v)P(\bm v')
&=
(-1)^{J(\bm v, \bm v')}P(\bm v')P(\bm v),
\label{eq:pauli_commutation}
\end{align}
where $J: \mathbb F_2^{2L} \times \mathbb F_2^{2L} \to \mathbb F_2$ is a bilinear form,
\begin{align}
J(\bm v, \bm v')
&=
\bm z \cdot \bm x'+\bm x \cdot \bm z'.
\label{eq:symplectic_form}
\end{align}
If we perform unitary evolution by $U$ on both sides of Eq.~\eqref{eq:pauli_commutation}, 
we obtain
\begin{align}
U P(\bm v)P(\bm v') U^\dagger
&=
\eta(\bm v)P(M\bm v)\eta(\bm v') P(M\bm v')
\notag
\\
&=
(-1)^{J(\bm v, \bm v')}\eta(\bm v') P(M\bm v') \eta(\bm v) P(M\bm v).
\end{align}
By comparing the resulting relation
with Eq.~\eqref{eq:pauli_commutation} evaluated at
$M\bm v$ and $M\bm v'$, we find
\begin{align}
J(M\bm v, M\bm v')
&=
J(\bm v, \bm v').
\label{eq:symplectic_invariance}
\end{align}

In matrix notation, the bilinear form is written as
\begin{align}
J(\bm u, \bm v)
&=
\bm u^T J \bm v
\end{align}
with
\begin{align}
J
&=
\begin{pmatrix}
0 & I_L \\
-I_L & 0
\end{pmatrix}
\in GL(2L, \mathbb F_2).
\end{align}
Here $I_L$ is the $L\times L$ identity matrix, and $-I_L=I_L$ in $\mathbb F_2$. 
The matrix $J$ can be regarded as the standard symplectic form, since it is alternating and nondegenerate,
i.e., $J(\bm v,\bm v)=0$ for every $\bm v$, and $J$ is invertible.
The relation~\eqref{eq:symplectic_invariance} is therefore equivalent to
\begin{align}
M^T J M
&=
J.
\label{eq:symplectic_matrix}
\end{align}
Hence $M\in\mathrm{Sp}(2L,\mathbb F_2)$ is a symplectic matrix for every choice of
the gate coefficients.
The additional restrictions on these coefficients needed
for eigenstate thermalization will be formulated through
the characteristic polynomial and the order of $M$.

\subsection{Characteristic polynomial}

To construct a Floquet circuit satisfying the strong ETH, we choose
the gate coefficients $\{a_i\}$ so that the corresponding label
transformation matrix $M$ has suitable spectral and orbit properties.
These properties are encoded in the characteristic polynomial,
\begin{align}
p_L(t)
&=
\det(tI-M),
\end{align}
which is defined over $\mathbb F_2$ (i.e., $p_L(t)\in \mathbb F_2[t]$).

To determine $p_L(t)$, it is convenient to consider
the ordered product,
\begin{align}
\widetilde M
&=
R_1(a_1)R_2(a_2) \cdots R_L(a_L).
\end{align}
For the circuit considered here, this ordering can be obtained
from Eq.~\eqref{eq:matrix_M} by commuting Clifford gates acting on
disjoint sites and cyclically permuting the gates.
Since cyclic permutations of Clifford gates can be expressed by
unitary evolution with Clifford unitaries,
$\widetilde M$ and $M$ are related through
a similarity transformation,
\begin{align}
\widetilde M
&=
CMC^{-1},
\end{align}
where $C$ is a symplectic matrix over $\mathbb F_2$
induced by a Clifford unitary.
Thus, $M$ and $\widetilde M$ share the same characteristic polynomial,
\begin{align}
p_L(t)
&=
\det(tI-\widetilde M).
\end{align}

To evaluate $\det(tI-\widetilde M)$, we perform Gaussian elimination over $\mathbb F_2[t]$.
First, we explicitly write down the formal equation
$(tI-\widetilde M)(\bm x,\bm z)^T=0$,
\begin{align}
t x_1+z_L &=0,\\
t x_i+x_{i-1}+a_{i-1}z_L &=0
\quad (2\le i\le L),\\
t z_1+x_L+\sum_{i=1}^{L}a_i z_i &=0,\\
t z_i+z_{i-1} &=0
\quad (2\le i\le L).
\end{align}
Here all coefficients are evaluated in characteristic 2.

Now we eliminate $x_1,\ldots,x_L$ using the first $L$
equations. Each elimination uses the coefficient $t$,
giving an overall factor $t^L$ in the determinant.
The resulting expression for $x_L$ is
\begin{align}
x_L
&=\left(
t^{-L}+\sum_{i=1}^{L-1}a_i t^{-(L-i)}
\right)z_L.
\end{align}
Next, we eliminate $z_1,\ldots,z_{L-1}$
using the last $L-1$ equations,
\begin{align}
z_i=t^{L-i}z_L
\quad (1\le i\le L).
\end{align}
These eliminations use coefficients equal to unity and
does not introduce additional factors in the determinant.
The remaining equation is
\begin{align}
\left[
t^L+\sum_{i=1}^{L}a_i t^{L-i}
+t^{-L}+\sum_{i=1}^{L-1}a_i t^{-(L-i)}
\right]z_L=0.
\end{align}
Multiplying the remaining coefficient by $t^L$, we obtain
\begin{align}
p_L(t)
&=\det(tI-\widetilde M)\notag\\
&=t^L\left[
t^L+\sum_{i=1}^{L}a_i t^{L-i}
+t^{-L}+\sum_{i=1}^{L-1}a_i t^{-(L-i)}
\right]\notag\\
&=t^{2L}+1+
\sum_{i=1}^{L-1}a_i(t^i+t^{2L-i})+a_Lt^L.
\end{align}
The bulk gate coefficients enter through pairs
of terms, $t^i$ and $t^{2L-i}$,
while the endpoint coefficient $a_L$ determines
the central term $t^L$.

\subsection{Gate coefficients}
\label{sec:gate_coefficients}

We choose the gate coefficients $\{a_j\}$ so that
$p_L(t)$ satisfies the following two conditions:
(i) it is irreducible over $\mathbb F_2$, and
(ii) its roots have the multiplicative order $D+1=2^L+1$.

The first condition (i) means that $p_L(t)$ cannot be factored
into two polynomials over $\mathbb F_2$.
In terms of the circuit dynamics, this ensures that
the Pauli-label space has no nonzero proper subspace
invariant under $M$.
Thus, the label dynamics cannot be decomposed into
independent sectors.
In particular, the evolution of any nonzero label
spans the full $2L$-dimensional label space.

The second condition (ii) fixes the period of the label dynamics.
A root $\zeta$ of $p_L(t)$ satisfies
\begin{align}
\zeta^{D+1}&=1,
&
\zeta^n&\ne1
\qquad (1\le n\le D).
\end{align}
Together with irreducibility, this implies
\begin{align}
M^{D+1}\bm v&=\bm v,
&
M^n\bm v&\ne\bm v
\qquad (1\le n\le D)
\label{eq:Singer_condition}
\end{align}
for every nonzero label $\bm v$.
An irreducible symplectic matrix
$M\in{\rm Sp}(2L,\mathbb F_2)$ of order $D+1$ (i.e., satisfying the relations (\ref{eq:Singer_condition}))
is called a Singer cycle.
From the viewpoint of the Floquet circuit, every
nontrivial Pauli string visits $D+1$ distinct labels
before returning to its initial label:
\begin{align}
\bm v
\to M\bm v
\to M^2\bm v
\to\cdots
\to M^{D+1}\bm v=\bm v.
\label{eq:Pauli_orbit}
\end{align}
Thus, we can classify the $D^2-1$ nonzero labels into
$(D^2-1)/(D+1)=D-1$ orbits of equal length.

The condition (i) fixes the endpoint coefficient $a_L$.
To see this, let us evaluate the characteristic polynomial at $t=1$,
\begin{align}
p_L(1)
&=
1+1+\sum_{j=1}^{L-1}a_j(1+1)+a_L
=
a_L,
\end{align}
where we have used $1+1=0$ in $\mathbb F_2$.
If $a_L=0$, then $t=1$ is a root of $p_L(t)$ and $p_L(t)$ can be factored as $(t-1)\times (\mbox{polynomial})$, making it reducible.
Irreducibility therefore requires $a_L=1$, which fixes
the endpoint gate as $G_L=SH$.

The irreducibility of $p_L(t)$ allows us to represent
the Pauli-label dynamics as multiplication in a finite field.
Since $p_L(t)$ has degree $2L$, we can define the finite field,
\begin{align}
\mathbb F_{D^2}
&\simeq
\mathbb F_2[t]/p_L(t).
\end{align}
Given that $p_L(t)$ is irreducible,
the root $\zeta$ of $p_L(t)$ can be regarded as an element of this field.
Every field element $v\in \mathbb F_{D^2}$ can be written uniquely as
\begin{align}
v
&=
\sum_{j=0}^{2L-1}c_j\zeta^j
\end{align}
with $c_j\in\mathbb F_2$.

\begin{table*}[t]
\caption{
Comparison of the Singer-cycle circuit and the LFSR circuit~\cite{IppolitiLong2026} with $D=2^L$. 
Periods and orbit counts refer to Pauli labels, with phases omitted.
In the LFSR circuit, the eigenvalue 1 of the Floquet operator is two-fold degenerate.
}
\label{tab:circuit_comparison}
\begin{tabular}{
p{0.34\textwidth}
p{0.36\textwidth}
p{0.27\textwidth}}
\hline\hline
Property & Singer-cycle circuit & LFSR circuit \\
\hline

Target space of matrix action
& Pauli-string labels
& Computational basis \\

Characteristic polynomial of the matrix
& $p_L(t)=t^{2L}+1+
\sum_{j=1}^{L-1}a_j(t^j+t^{2L-j})+t^L$
& $g_L(t)=t^L+1+\sum_{j=1}^{L-1}b_j t^j$ \\

& : irreducible polynomial of degree $2L$
& : primitive polynomial of degree $L$ \\

Period of nonzero Pauli labels
& $D+1$
& $D-1$ \\

Number of nonzero Pauli-label orbits
& $D-1$
& $D+1$ \\

Eigenvalues of the Floquet operator
& $e^{2\pi ik/(D+1)}$ ($k=1,2,\dots,D)$
& $e^{2\pi ik/(D-1)}$ ($k=0,1,\dots,D-2$) \\

Nonthermal eigenstates
& None
& ${\rm Span}\{|0\rangle^{\otimes L}, |+\rangle^{\otimes L}\}$ \\

\hline\hline
\end{tabular}
\end{table*}

To identify the Pauli labels with these field elements,
let us choose a nonzero reference label
$\bm v_0\in\mathbb F_2^{2L}$.
Since $p_L(t)$ is irreducible, we can use the vectors
\begin{align}
\bm v_0,\;M\bm v_0,\;M^2\bm v_0,\;\dots,\;M^{2L-1}\bm v_0
\end{align}
as a basis of the label space $\mathbb F_2^{2L}$.
Thus, every label can be uniquely expanded as
\begin{align}
\bm v
&=
\sum_{j=0}^{2L-1}c_jM^j\bm v_0.
\end{align}
We define the correspondence between Pauli labels in $\mathbb F_2^{2L}$ and field elements in $\mathbb F_{D^2}$ as
\begin{align}
f:\quad
\bm v=\sum_{j=0}^{2L-1}c_jM^j\bm v_0
&\longmapsto
v=\sum_{j=0}^{2L-1}c_j\zeta^j.
\label{eq:map_f}
\end{align}
This map is one-to-one between the Pauli-label space $\mathbb F_2^{2L}$ and the finite field $\mathbb F_{D^2}$.
We can now express the action of $M$ as multiplication by $\zeta$:
\begin{align}
f(M\bm v)
&=
\zeta f(\bm v).
\end{align}
The map $f$ is consistent in the sense that 
the Cayley--Hamilton relation $p_L(M)=0$ and the root relation $p_L(\zeta)=0$
reduce higher
powers of $M$ and $\zeta$ in exactly the same way.
Consequently, the Pauli label $v$ is mapped to $\zeta^n v$ after $n$ periods
in the Floquet circuit,
with the accumulated phase tracked separately.

One can see that $D+1$ is the largest possible multiplicative order of a root of $p_L(t)$.
This can be understood as follows. First, we regard a root $\zeta$ of $p_L(t)$ as an element of the multiplicative group
$\mathbb F_{D^2}^\times$.
Since this group contains $D^2-1$ elements,
Lagrange's theorem gives
\begin{align}
\zeta^{D^2-1}
&=
1.
\end{align}
This provides a bound on the order of $\zeta$, which can be sharpened by using the form of $p_L(t)$.

If $\zeta$ is the root of $p_L(t)$, then so is $\zeta^2$, since
\begin{align}
p_L(\zeta^2)
&=
p_L(\zeta)^2
=
0,
\end{align}
where we have used $a_j^2=a_j (\in\mathbb F_2)$.
By repeatedly taking squares (Frobenius map), one finds that
\begin{align}
\zeta, \; \zeta^2, \; (\zeta^2)^2, \; \cdots, \; \zeta^{2^{2L-1}}
\label{eq:roots_p_L}
\end{align}
are all the roots of $p_L(t)$ (note that $\zeta^{2^{2L}}=\zeta$). 
Since $p_L(t)$ is irreducible, all of these roots are distinct from each other:
If $\zeta^{2^j}=\zeta$ for some $1\le j<2L$,
then $\zeta$ would belong to $\mathbb F_{2^j}$,
and its minimal polynomial over $\mathbb F_2$
would have degree at most $j$.
This would contradict the irreducibility of $p_L(t)$.
Thus, the roots of $p_L(t)$ are given by (\ref{eq:roots_p_L}).

We also notice that the polynomial $p_L(t)$ satisfies the reciprocity relation,
\begin{align}
p_L(t)
&=
t^{2L} p_L(t^{-1}),
\label{eq:reciprocity_p_L}
\end{align}
which implies that $\zeta^{-1}$ is also a root.
We can therefore write
$\zeta^{-1}=\zeta^{2^j}$ for some $0\le j<2L$.
By repeatedly using this relation, we obtain
\begin{align}
\zeta^{2^{2j}}
&=
(\zeta^{2^j})^{2^j}
=
(\zeta^{-1})^{2^j}
=
(\zeta^{2^j})^{-1}
=
\zeta.
\end{align}
This means $2j=2L$,
i.e., $j=L$, leading to
\begin{align}
\zeta^{2^L}
&=
\zeta^{-1}
\quad
\Leftrightarrow
\quad
\zeta^{D+1}
=
1.
\end{align}
Thus, the multiplicative order of $\zeta$ has a maximum bound of $D+1$.

We can explicitly construct a characteristic polynomial satisfying
both of the conditions (i) and (ii) by choosing the gate coefficients $\{a_j\}$ appropriately.
To this end, we use the fact that $\mathbb F_{D^2}^{\times}$
is a cyclic group of order $D^2-1=(D-1)(D+1)$.
By denoting a generator of $\mathbb F_{D^2}^{\times}$ by $g$, we can write
\begin{align}
\mathbb F_{D^2}^\times
&=
\{g, g^2, g^3, \dots, g^{D^2-1}=1\}.
\end{align}
If we choose $\zeta=g^{D-1}$, then
\begin{align}
\zeta^{D+1}
&=
(g^{D-1})^{D+1}
=
g^{D^2-1}
=
1.
\end{align}
Since $g$ has order $D^2-1$, $\zeta$ has order
exactly $D+1$.
We note that $\zeta, \zeta^2, \dots, \zeta^{2^{2L-1}}$ are all different.
If $\zeta^{2^j}=\zeta$ for some $1\le j<2L$, then
$\zeta^{2^{2L-j}}=\zeta$ also follows from $(\zeta^{2^j})^{2^{2L-j}}=\zeta^{2^{2L-j}}$.
Thus, setting $k=\min(j,2L-j)$, we obtain
$\zeta^{2^k-1}=1$ with $1\le k\le L$.
This contradicts the fact that $\zeta$ has order $2^L+1$, since $2^k-1<2^L+1$.

\begin{table*}[t]
\setlength{\tabcolsep}{28pt}
\caption{
Examples of binary gate coefficients in the Singer-cycle circuit and the LFSR circuit~\cite{IppolitiLong2026}
for the system size $L=2, 3, \dots, 18$.
The examples of the coefficients $b_1b_2\cdots b_{L-1}$ for $8\le L\le 18$ are the same as those in Ref.~\cite{IppolitiLong2026}
(for $L=14$ we show a corrected example).
Periods refer to nonzero Pauli labels, with phases omitted.
}
\label{tab:clifford_coefficients}
\begin{tabular}{clrlr}
\hline\hline
\toprule
& \multicolumn{2}{c}{Singer-cycle circuit}
& \multicolumn{2}{c}{LFSR circuit} \\
\cmidrule(lr){2-3}\cmidrule(lr){4-5}
$L$ & $a_1a_2\cdots a_L$ & Period
& $b_1b_2\cdots b_{L-1}$ & Period \\
\hline
\midrule
2 & \texttt{11} & 5 & \texttt{1} & 3 \\
3 & \texttt{001} & 9 & \texttt{10} & 7 \\
4 & \texttt{0011} & 17 & \texttt{100} & 15 \\
5 & \texttt{10001} & 33 & \texttt{0100} & 31 \\
6 & \texttt{010011} & 65 & \texttt{10000} & 63 \\
7 & \texttt{0000101} & 129 & \texttt{100000} & 127 \\
8 & \texttt{10000001} & 257 & \texttt{0111000} & 255 \\
9 & \texttt{010000011} & 513 & \texttt{00010000} & 511 \\
10 & \texttt{0010000001} & 1025 & \texttt{001000000} & 1023 \\
11 & \texttt{10001000001} & 2049 & \texttt{0100000000} & 2047 \\
12 & \texttt{100000010001} & 4097 & \texttt{11000001000} & 4095 \\
13 & \texttt{0000000000101} & 8193 & \texttt{110010000000} & 8191 \\
14 & \texttt{01000010000001} & 16385 & \texttt{1100000000010} & 16383 \\
15 & \texttt{000000100000101} & 32769 & \texttt{10000000000000} & 32767 \\
16 & \texttt{0000000000100001} & 65537 & \texttt{101000000001000} & 65535 \\
17 & \texttt{10001000000000001} & 131073 & \texttt{0010000000000000} & 131071 \\
18 & \texttt{110000000000000001} & 262145 & \texttt{00000010000000000} & 262143 \\
\bottomrule
\hline\hline
\end{tabular}
\end{table*}

Let us then consider a polynomial over $\mathbb F_{D^2}$ defined by
\begin{align}
q_L(t)
&=
(t-\zeta)(t-\zeta^2) \cdots (t-\zeta^{2^{2L-1}}).
\end{align}
If we write $q_L(t)=\sum_{j=0}^{2L} c_j t^j$,
the coefficients $c_j$ satisfy $c_j^2=c_j$ (i.e., $c_j=0, 1$), since $q_L(t)^2=q_L(t^2)$ 
over $\mathbb F_{D^2}$ with characteristic 2.
Thus, $q_L(t)$ belongs to $\mathbb F_2[t]$.
Moreover, any polynomial over $\mathbb F_2$ that
vanishes at $\zeta$ must also vanish at all the
$2L$ distinct roots, $\zeta, \zeta^2, \dots, \zeta^{2^{2L-1}}$.
We therefore identify $q_L(t)$ as the minimal polynomial having the root $\zeta$. 
Since the minimal polynomial is irreducible, $q_L(t)$ is irreducible as well.
If we replace each root by its inverse, the roots are permuted due to the relation $\zeta^{-1}=\zeta^{2^L}$.
Together with $c_0=c_{2L}=1$, this gives
$c_j=c_{2L-j}$.
Thus, $q_L(t)$ has the same reciprocal form
as $p_L(t)$ [Eq.~(\ref{eq:reciprocity_p_L})].
Hence we prove the existence of the polynomial $p_L(t)$ satisfying
the two conditions (i) and (ii) for every $L$.

So far, we showed a concrete example of $p_L(t)$ with the root satisfying $\zeta=g^{D-1}$ 
($g$ is a generator of $\mathbb F_{D^2}^\times$).
More generally, we can show that 
there always exists $g\in \mathbb F_{D^2}$ such that $g$ becomes a generator of $\mathbb F_{D^2}^\times$
and satisfies $g^{D-1}=\zeta$ for a given $p_L(t)$. 
Let us denote a generator of $\mathbb F_D^\times$ by $h$ (which satisfies $h^{D-1}=1$).
If we put $g=h\zeta^{D/2}$, then we obtain
\begin{align}
g^{D-1}
&=
h^{D-1}\zeta^{D(D-1)/2}
\notag
\\
&=
\zeta^{(D+1-2)D/2}
=
\zeta^{-D}
=
\zeta.
\end{align}
Since $D+1$ and $D/2$ are coprime, the order of $\zeta^{D/2}$ is $D+1$.
Since $D+1$ and $D-1$ are also coprime, the order of $g=h\zeta^{D/2}$ is given by
$(D-1)(D+1)=D^2-1$, which implies that $g$ is a generator of $\mathbb F_{D^2}^\times$.

In Table~\ref{tab:circuit_comparison}, we compare the properties of the Singer-cycle circuit
with those of the LFSR circuit.
Both constructions use characteristic polynomials over $\mathbb F_2$
to control the circuit dynamics, but the corresponding matrices act on different spaces:
In the Singer-cycle circuit, the matrix $M$ acts on Pauli string labels $\bm v$,
while in the LFSR circuit the matrix $A$ acts on the computational basis of $L$ qubits.
In the latter, one chooses a primitive polynomial $g_L(t)$ of degree $L$ to generate a single orbit containing all
$D-1$ nonzero computational-basis bitstrings.
The circuit preserves the pure-$X$ and pure-$Z$ Pauli
sectors separately and admits the product eigenstates
$|0\rangle^{\otimes L}$ and $|+\rangle^{\otimes L}$,
which violate the strong ETH.
In our circuit, we instead mix the $X$ and $Z$ sectors
through the endpoint gate $SH$ and use an irreducible reciprocal polynomial of degree $2L$.
Its roots have order $D+1$, so that the polynomial $p_L(t)$ is not primitive
for $L\ge2$.
With this choice, we eliminate all nonzero proper
invariant subspaces of the Pauli-label space.
In particular, no Floquet eigenstate can be a stabilizer
state, since its stabilizer labels would span an invariant
$L$-dimensional subspace.
Although both circuits exhibit exponentially long
Pauli-label periods, $D-1$ and $D+1$, respectively,
the key distinction is therefore how they mix the
Pauli sectors and constrain the eigenstates.

In Table~\ref{tab:clifford_coefficients}, we show examples of binary coefficients
for the Singer-cycle circuit and the LFSR circuit.
For our circuit, we fix $a_L=1$ and perform brute-force search over the remaining
binary coefficients $a_1,\ldots,a_{L-1}$
in increasing order of the number of bulk CNOT gates.
For each candidate, we examine the irreducibility of $p_L(t)$ over $\mathbb F_2$
and the multiplicative order of the roots.
The resulting examples therefore minimize the number
of bulk CNOT gates at each size.
One can see that the examples in Table~\ref{tab:clifford_coefficients}
require remarkably few CNOT gates.
For the present circuit, we need at most two bulk CNOT gates
for $L\le 18$, with no apparent growth as $L$ increases.
This observation suggests that sparse gate configurations
would suffice even for larger systems. 
In general, we do not find a simple periodic pattern
in the coefficients, either along the chain or as
the system size increases.
For the LFSR circuit, we show examples of the coefficients $b_1, b_2, \dots, b_{L-1}$.
For $8\le L\le 18$, they are the same as those listed in Ref.~\cite{IppolitiLong2026} (for $L=14$ we show a corrected example).
From these examples, one can see that the sequence of coefficients $a_1, \dots, a_{L-1}$ ($a_L=1$)
is generally different from $b_1, \dots, b_{L-1}$.

\subsection{Eigenvalues and eigenstates of the Floquet operator}
\label{sec:eigenvalues_U}

Having established the conditions on the gate coefficients, 
we now study the eigenvalues and eigenstates of the Floquet operator $U$ \eqref{eq:brickwork_matrix}, 
whose action on Pauli labels is described by Eq.~\eqref{eq:clifford_action}.
We denote the period of nonzero Pauli labels as
\begin{align}
N
&=
D+1.
\end{align}
Since $M^N=I$, every nonzero Pauli label returns to its initial label
after $N$ cycles, 
\begin{align}
U^N P(\bm v) U^{-N}
&=
\eta_N(\bm v) P(\bm v),
\end{align}
where $\eta_N(\bm v)$ is the accumulated phase.

Let us first show that $\eta_N(\bm v)=1$.
Since $P(\bm v)P(\bm v')\propto P(\bm v+\bm v')$, we notice that
\begin{align}
\eta_N(\bm v+\bm v')
&=
\eta_N(\bm v)\eta_N(\bm v'),
\end{align}
which implies $\eta_N(\bm v)^2=1$.
Because $U$ commutes with $U^N$, we have
$\eta_N(M\bm v)=\eta_N(\bm v)$.
Combining these relations, we obtain
\begin{align}
\eta_N((M-I)\bm v)
&=
\eta_N(M\bm v)\eta_N(\bm v)
=
\eta_N(\bm v)^2
=
1.
\end{align}
Since $M$ does not have an eigenvalue 1, the matrix $M-I$ is invertible.
Thus, $\eta_N(\bm v)=1$ for every label $\bm v$.

As $U^N$ commutes with every Pauli string
and the Pauli strings span the full operator space,
we conclude that
\begin{align}
U^N
&=
e^{i\theta}I
\end{align}
with $\theta\in\mathbb R$.
Thus, every quantum state returns to itself
after $N$ periods up to a common phase.
By redefining the overall phase of $U$, we may set
\begin{align}
U^N
&=
I.
\end{align}
The eigenvalues of $U$ then belong to the set of $N$th roots
of unity,
\begin{align}
U|\psi_k\rangle
&=
e^{2\pi i k/N}|\psi_k\rangle,
\end{align}
with $k\in \{0,1,2,\dots, N-1\}$ ($|\psi_k\rangle$ are the corresponding eigenstates).
This restricts the possible eigenvalues. 

Next we show that the eigenvalues of $U$ are nondegenerate
and, after a suitable choice of the overall phase, are given by
\begin{align}
\omega^k
&=
e^{2\pi i k/N}
\quad
(k=1,2,\dots,N-1).
\label{eq:eigenvalue_U}
\end{align}
To this end, we count the number of linearly independent operators that commute with $U$.
For a given operator $Q$ that commutes with $U$, we expand it in the Pauli-operator basis as $Q=\sum_{\bm v} c_{\bm v} P(\bm v)$. From $UQU^\dagger=Q$, we find that $c_{M\bm v}=\eta(\bm v)c_{\bm v}$.
Thus, the coefficients are uniquely determined from a single coefficient in each orbit of Pauli labels \eqref{eq:Pauli_orbit}, and there is at most one linearly independent commuting operator in each orbit.
In fact, we can explicitly construct such an operator as $\sum_{j=0}^{N-1} U^j P(\bm v_0) U^{-j}$, where $\bm v_0$ is an initial nonzero label of the orbit. Since there are $D-1$ orbits, together with the identity operator $I$ we find that the number of independent operators that commute with $U$ is exactly $D$.

If we denote multiplicities of the eigenvalues of $U$ by $m_k$, 
the number of linearly independent operators commuting
with $U$ is given by $\sum_k m_k^2$.
We therefore have
\begin{align}
\sum_k m_k^2
&=
D
=
\sum_k m_k,
\end{align}
which implies that every eigenvalue is nondegenerate (i.e., $m_k=1$).
Since $D=N-1$, exactly one of the $N$th roots of unity
is absent in the set of the eigenvalues.
We choose the overall phase of $U$ so that the missing
root is unity (this choice is consistent with $U^N=1$).
The resulting eigenvalues are given by Eq.~\eqref{eq:eigenvalue_U}.
One can see that the eigenvalues of $U$ distribute with equal spacing 
(except for the one between $e^{2\pi i/N}$ and $e^{2\pi i(N-1)/N}$) 
along the unit circle in the complex plane.

The eigenstates of $U$ can be labeled by the characters of
the finite cyclic group generated by $\zeta$ (a root of $p_L(t)$),
\begin{align}
T
&=
\{1,\zeta,\zeta^2,\dots,\zeta^{N-1}\},
\label{eq:group_T}
\end{align}
which is a subgroup of $\mathbb F_{D^2}^{\times}$.
Since $U^N=I$, we obtain a unitary representation of $T$
by associating $\zeta^n$ with $U^n$.
For each eigenstate $|\psi_k\rangle$, we define a character
$\chi_k:T\to\mathbb C$ by
\begin{align}
\chi_k(\zeta^n)
&=
\omega^{kn}
=
e^{2\pi i kn/N}
\quad
(k=0,1,\dots,N-1).
\end{align}
These characters satisfy
$\chi_k(r)\chi_k(s)=\chi_k(rs)$ for $r,s\in T$,
$\chi_k(s)\chi_{k'}(s)=\chi_{k+k'}(s)$ for $k, k'=0,1,\dots,N-1$ (mod $N$),
and $\overline{\chi_k(s)}=\chi_{-k}(s)$ (here the overline denotes complex conjugation).
The characters obey the orthogonality relation:
\begin{align}
\frac{1}{N}
\sum_{s\in T} \overline{\chi_k(s)}\chi_{k'}(s)
&=
\delta_{k,k'}.
\label{eq:orthogonality_chi}
\end{align}

Using the characters, we can describe how the corresponding eigenstates evolve under $U$:
\begin{align}
U|\psi_k\rangle
&=
\chi_k(\zeta)|\psi_k\rangle.
\end{align}
With our choice of the overall phase of $U$, every
nontrivial character occurs exactly once, while the trivial
character $\chi_0(r)=1$ is absent.
We therefore write the eigenstates of $U$ as $|\chi\rangle$,
with $\chi$ running over the nontrivial characters of $T$.

For each $s=\zeta^n\in T$, we write
\begin{align}
U_s
&=
U^n.
\end{align}
Since $U^N=I$, this definition does not depend on the
choice of the integer representative $n$ modulo $N$.
The operators $U_s$ satisfy $U_r U_s=U_{rs}$ and $U_1=I$,
and thus form a unitary representation of $T$.
In this notation, the eigenstates obey
\begin{align}
U_s|\chi\rangle
&=
\chi(s)|\chi\rangle
\quad (s\in T).
\end{align}
Using $\sum_{k=1}^{N-1} \omega^{kn}=\sum_{k=0}^{N-1} \omega^{kn}-1=N\delta_{n, 0}-1$,
one can quickly confirm that the trace of $U_s$ is given by
\begin{align}
{\rm Tr}\, U_s
&=
\begin{cases}
D & (s=1) \\
-1 & (s\neq 1)
\end{cases}.
\label{eq:Tr_U_s}
\end{align}

\subsection{Phase convention over the finite field}

To establish ETH, we need to evaluate matrix elements of observables in the eigenbasis of $U$.
Our strategy is to express relevant quantities entering these matrix elements in the language of finite fields.
We already expressed the eigenvalues and eigenstates of $U$ in terms of finite fields (see Sec.~\ref{sec:eigenvalues_U}). 
We also identified Pauli labels $\bm v$
and elements $v$ of the finite field $\mathbb F_{D^2}$ (Sec.~\ref{sec:gate_coefficients}).
In what follows, we write $P(v)$ for a Pauli operator $P(\bm v)$ with the label $v=f(\bm v)$.

What remains is to express phase factors that appear when manipulating Pauli operators in terms of finite fields.
When we take a product of Pauli operators and compare it with a Pauli operator with the summed label, we obtain a phase factor,
\begin{align}
P(v)P(v')
&=
(-1)^{B_0(v, v')}
P(v+v'),
\label{eq:P_product}
\end{align}
where $B_0: \mathbb F_{D^2} \times \mathbb F_{D^2} \to \mathbb F_2$ is a bilinear form.
If we write $v=f(\bm x, \bm z)$ and $v'=f(\bm x', \bm z')$, the phase exponent is explicitly given by
\begin{align}
B_0(v, v')
&=
\bm z \cdot \bm x'.
\end{align}
Although $B_0$ is not symmetric, it reproduces the
symplectic form (\ref{eq:symplectic_form}) through
\begin{align}
B_0(v, v')+B_0(v', v)
&=
J(v, v').
\end{align}

If we take a Hermitian conjugate of Pauli operators, we obtain another phase factor,
\begin{align}
P(v)^\dagger
&=
(-1)^{Q_0(v)}P(v),
\label{eq:P_dagger}
\end{align}
where $Q_0: \mathbb F_{D^2}\to \mathbb F_2$ is defined by
\begin{align}
Q_0(v)
&=
B_0(v, v)
=
\bm z \cdot \bm x.
\end{align}
Thus, $P(v)$ is Hermitian when $Q_0(v)=0$ and
anti-Hermitian when $Q_0(v)=1$.

For later calculations, we need to evaluate the sum of the phase exponents
over a full period of nonzero Pauli labels.
Since the label $v$ evolves as $v\mapsto\zeta v$,
we define
\begin{align}
B(v,v')
&=
\sum_{j=0}^{N-1}B_0(\zeta^jv,\zeta^jv')
=
\sum_{s\in T}B_0(sv,sv'),
\label{eq:B_def}
\end{align}
and
\begin{align}
Q(v)
&=
B(v,v)
=
\sum_{s\in T}Q_0(sv).
\label{eq:Q_def}
\end{align}
We evaluate these sums in $\mathbb F_2$, so that
$(-1)^{B(v,v')}$ and $(-1)^{Q(v)}$ give the products
of the corresponding sign factors over one period.
Since multiplying all labels by any $r\in T$ only permutes the terms in the sums,
we have
\begin{align}
B(rv,rv')
&=
B(v,v'),
\label{eq:B_r_invariance}
\\
Q(rv)
&=
Q(v),
\end{align}
for $r\in T$.
These identities allow us to work with phase functions
that are invariant under the Clifford label dynamics.

We further show that the phase exponents $B(v,v')$ and $Q(v)$ can be expressed
in terms of finite-field traces.
Let $K=\mathbb F_D$ and $E=\mathbb F_{D^2}$, 
where $D=2^L$.
We define the finite-field traces by
\begin{align}
\operatorname{tr}_K(x)
&=
\sum_{j=0}^{L-1}x^{2^j}
\in\mathbb F_2
\quad (x\in K),
\\
\operatorname{tr}_E(v)
&=
\sum_{j=0}^{2L-1}v^{2^j}
\in\mathbb F_2
\quad (v\in E).
\end{align}
These maps are called `traces' because they coincide with
ordinary matrix traces in the following sense.
If one views $K$ as an $L$-dimensional vector
space over $\mathbb F_2$, multiplication by $x\in K$
can be represented by an $L\times L$ matrix $M_x$ over $\mathbb F_2$.
Since the eigenvalues of $M_x$ (over $K$) are given by $x, x^2, \dots, x^{2^{L-1}}$,
the trace of $M_x$ is precisely $\operatorname{tr}_K(x)$.
The same interpretation applies to $\operatorname{tr}_E(v)$,
with $E$ viewed as a $2L$-dimensional vector space
over $\mathbb F_2$.
These traces are additive:
\begin{align}
{\rm tr}_K(x+y)
&=
{\rm tr}_K(x)+{\rm tr}_K(y)
\quad
(x,y\in K),
\\
{\rm tr}_E(u+v)
&=
{\rm tr}_E(u)+{\rm tr}_E(v)
\quad
(u,v\in E),
\end{align}
which follows from
$(a+b)^{2^j}=a^{2^j}+b^{2^j}$ in characteristic 2.

We also define conjugation on $E$ by
\begin{align}
\bar v
&:=
v^D
\quad
(v\in E),
\end{align}
which is a finite-field analog of complex conjugation
since $\overline{u+v}=\bar u+\bar v$,
$\overline{uv}=\bar u\,\bar v$,
and $\overline{\bar v}=v$.
Note that elements satisfying
$\bar x=x$ are precisely the $D$ roots of $t^D-t$,
and hence $\bar x=x \Leftrightarrow x\in K$.
In particular, $v+\bar v\in K$ for every $v\in E$,
because $\overline{v+\bar v}=v+\bar v$.
The traces on $E$ and $K$ are related by
\begin{align}
\operatorname{tr}_E(v)
&=
\operatorname{tr}_K(v+\bar v)
\quad (v\in E),
\end{align}
as follows directly from
\begin{align}
\operatorname{tr}_K(v+\bar v)
&=
\sum_{j=0}^{L-1}
\left(v^{2^j}+v^{2^{L+j}}\right)
=
\operatorname{tr}_E(v).
\end{align}
Conjugation preserves the trace over $E$, since ${\rm tr}_E(\bar v)=
{\rm tr}_K(\bar v+\overline{\bar v})={\rm tr}_K(\bar v+v)={\rm tr}_E(v)$.

To express $B(v, v')$ and $Q(v)$ with the finite-field traces,
we first fix the normalization of the finite-field coordinates.
Since the trace pairing $(v,v')\mapsto{\rm tr}_E(vv')$ is nondegenerate
(i.e., if ${\rm tr}_E(vv')=0$ for all $v'\in E$, then $v=0$),
the symplectic form $J(v,v')$
in Eq.~\eqref{eq:symplectic_form} can be written uniquely as
\begin{align}
J(v, v')={\rm tr}_E(v A_0(v')),
\end{align}
where $A_0: E\to E$ is an $\mathbb F_2$-linear map.
From the relation $J(\zeta v, \zeta v')=J(v, v')$ [Eq.~(\ref{eq:symplectic_invariance})]
together with the nondegeneracy of the trace pairing,
we find that $A_0(\zeta v)=\zeta^{-1}A_0(v)$.
If we put $\lambda:=A_0(1)\in E$ and expand $v\in E$ as $v=\sum_{j=0}^{2L-1} c_j \zeta^j$
($c_j\in \mathbb F_2$), we obtain
\begin{align}
A_0(v)
&=
\sum_{j=0}^{2L-1} c_j A_0(\zeta^j)
=
\sum_{j=0}^{2L-1} c_j \zeta^{-j} \lambda
\notag
\\
&=
\lambda \sum_{j=0}^{2L-1} c_j \bar\zeta^j
=
\lambda \bar v,
\label{eq:A_0}
\end{align}
where we have used $\bar\zeta=\zeta^D=\zeta^{-1}$.
Using Eq.~\eqref{eq:A_0}, we can rewrite the symplectic form as
\begin{align}
J(v, v')
&=
{\rm tr}_E(\lambda v\overline{v'}).
\label{eq:J_tr_E}
\end{align}

Due to the symmetry of $J$ ($J(v,v')=J(v',v)$), we find that $\lambda=\bar\lambda\, (\in K)$.
To show this, let us note that the trace is invariant under conjugation,
\begin{align}
J(v',v)
&=
\operatorname{tr}_E(\lambda v'\bar v)
=
\operatorname{tr}_E(\bar\lambda v\overline{v'}).
\end{align}
Comparing this with Eq.~\eqref{eq:J_tr_E}, we obtain
\begin{align}
\operatorname{tr}_E
\bigl((\lambda+\bar\lambda)v\overline{v'}\bigr)
&=
0
\quad
(v,v'\in E).
\end{align}
If we set $v'=1$ and use the nondegeneracy of the trace pairing, we conclude that $\lambda=\bar\lambda$,
and hence $\lambda\in K$.

Now we rescale the map $f: \mathbb F_2^{2L}\to \mathbb F_{D^2}$ \eqref{eq:map_f}
between the Pauli-label space and the finite field as $f\mapsto \tilde f=\alpha f$
with $\alpha\in E$ ($\alpha\neq 0$). This is always possible without changing the relation
$f(M\bm v)=\zeta f(\bm v)$. In other words, we have a kind of gauge degrees of freedom to define the map $f$.
We particularly choose $\alpha=\lambda^{D/2}$,
which satisfies $\bar\alpha=\bar\lambda^{D/2}=\lambda^{D/2}=\alpha$
and $\alpha\bar\alpha=\alpha^2=\lambda^D=\lambda$ (since $\lambda\in K$).
Thus, the symplectic form becomes
\begin{align}
J(\bm v, \bm v')
&=
{\rm tr}_E(\lambda f(\bm v) \overline{f(\bm v')})
=
{\rm tr}_E(\tilde f(\bm v) \overline{\tilde f(\bm v')}).
\end{align}
If we use the rescaled coordinates defined by $\tilde f$, we obtain
\begin{align}
J(v, v')
&=
{\rm tr}_E(v \overline{v'})
\label{eq:J_trace}
\end{align}
In the following, we use these rescaled coordinates for the finite field.
Note that the condition (\ref{eq:J_trace}) does not fully fix the gauge of $f$, but
leaves a freedom to rescale $f$ by $\zeta^n$.

We can apply a similar argument to the bilinear form $B(v, v')$ (\ref{eq:B_def}).
Since $B(v, v')$ satisfies $B(\zeta v, \zeta v')=B(v, v')$ [Eq.~(\ref{eq:B_r_invariance})], we can write
\begin{align}
B(v, v')
&=
{\rm tr}_E(\beta v \overline{v'})
\label{eq:B_trace}
\end{align}
for a unique $\beta\in E$.
Since $B_0(v, v')+B_0(v', v)=J(v, v')$,
we have
\begin{align}
B(v, v')+B(v', v)
&=
\sum_{s\in T} (B_0(sv, sv')+B_0(sv', sv))
\notag
\\
&=
\sum_{s\in T} J(sv, sv')
=
NJ(v, v')
\notag
\\
&=
J(v, v'),
\end{align}
where the last equality follows since $N=D+1$ is odd.
On the other hand, Eq.~\eqref{eq:B_trace} yields
\begin{align}
B(v,v')+B(v',v)
&=
\operatorname{tr}_E
\bigl((\beta+\bar\beta)v\overline{v'}\bigr).
\end{align}
Comparing this expression with Eq.~\eqref{eq:J_trace}
and using the nondegeneracy of the trace pairing,
we obtain
\begin{align}
\beta+\bar\beta
&=
1.
\label{eq:beta}
\end{align}
Later, we will see that the explicit value of $\beta$ is not needed
but Eq.~\eqref{eq:beta} suffices for the analysis.

Finally, since $v\bar v\in K$, we find
\begin{align}
Q(v)
&=
B(v,v)
=
\operatorname{tr}_E(\beta v\bar v)
\notag\\
&=
\operatorname{tr}_K
\bigl((\beta+\bar\beta)v\bar v\bigr)
=
\operatorname{tr}_K(v\bar v).
\label{eq:Q_trace}
\end{align}
Thus, in the normalized field coordinates with $J(v,v')={\rm tr}_E(v\overline{v'})$ [Eq.~\eqref{eq:J_trace}], 
$B(v,v')$ and $Q(v)$ are written in terms of the finite-field traces through Eqs.~\eqref{eq:B_trace} and \eqref{eq:Q_trace},
respectively.

\section{Strong eigenstate thermalization}
\label{sec:strong_ETH}

In this section, we establish the strong ETH \eqref{eq:infinite_T_ETH} for the Singer-cycle Clifford circuits. We first express the squared matrix elements
of Pauli operators as double sums over finite-field elements, and then reduce them to single sums of mixed characters. Applying the Weil bound \cite{Weil1948, Castro2000, Cochrane2006} to these sums yields uniform bounds on the matrix elements, from which the strong ETH follows for arbitrary observables supported on a fixed number of sites.

\subsection{Evaluation of squared matrix elements}

To prove the strong ETH in the present Singer-cycle circuit, we evaluate the matrix elements of
arbitrary non-identity Pauli strings with respect to the eigenstates of the Floquet operator.
Unlike the case of the LFSR circuit reviewed in Sec.~\ref{sec:LFSR_circuit}, 
the present Singer-cycle circuit
contains a Hadamard gate, which maps computational-basis
states to their superpositions.
Therefore, instead of explicitly constructing the eigenstates in the computational basis,
we consider the squared magnitudes of matrix elements,
\begin{align}
|\langle \chi | P(v)| \chi' \rangle|^2
&=
{\rm Tr}(\Pi_\chi P(v)\Pi_{\chi'}P(v)^\dagger)
\quad
(v\neq 0),
\label{eq:squared_matrix_elements}
\end{align}
where $\Pi_\chi=|\chi\rangle \langle\chi|$ is the projection operator onto the eigenstate $|\chi\rangle$.
The squared matrix elements \eqref{eq:squared_matrix_elements} can be evaluated without explicitly
representing the eigenstates in the computational basis.

Using the orthogonality [Eq.~\eqref{eq:orthogonality_chi}] of the characters $\chi$ of the cyclic group $T$ (\ref{eq:group_T}), we obtain
\begin{align}
\frac{1}{N}\sum_{s\in T} \overline{\chi(s)}U_s |\chi'\rangle
&=
\frac{1}{N}\sum_{s\in T} \overline{\chi(s)}\chi'(s) |\chi'\rangle
=
\delta_{\chi, \chi'}|\chi'\rangle.
\end{align}
Thus, the projection operator can be expressed as
\begin{align}
\Pi_\chi
&=
\frac{1}{N}\sum_{s\in T} \overline{\chi(s)}U_s.
\end{align}
Substituting this into Eq.~\eqref{eq:squared_matrix_elements}, we can write
\begin{align}
|\langle \chi | P(v)| \chi' \rangle|^2
&=
\frac{1}{N^2}\sum_{s,t\in T}
\overline{\chi(s)}\, \overline{\chi'(t)}
{\rm Tr}(U_s P(v) U_t P(v)^\dagger).
\end{align}
Let us denote the trace part of the above expression by $K_v(s,t)$:
\begin{align}
K_v(s,t)
&=
{\rm Tr}(U_s P(v) U_t P(v)^\dagger).
\end{align}
The rest of this subsection is devoted to evaluating the function $K_v(s,t)$ explicitly. 

To this end, let us first recall that a Pauli string $P(v)$ acquires a phase $\eta(v)$ after unitary evolution with $U$
[Eq.~(\ref{eq:clifford_action})]. After $n$ cycles of unitary evolution with $U^n=U_t$ ($t=\zeta^n$), 
the Pauli string transforms as
\begin{align}
U_t P(v) U_t^\dagger
&=
\eta_t(v) P(tv),
\label{eq:U_t_P}
\end{align}
where the accumulated phase is denoted by $\eta_t(v)=\eta(v)\eta(\zeta v)\cdots\eta(\zeta^{n-1}v)$.
Using Eq.~(\ref{eq:U_t_P}) together with the relations (\ref{eq:P_product}) and (\ref{eq:P_dagger}),
we rewrite $K_v(s,t)$ as
\begin{align}
K_v(s,t)
&=
{\rm Tr}(U_s U_t \eta_{t^{-1}}(v)P(t^{-1}v) (-1)^{Q_0(v)}P(v))
\notag
\\
&=
\eta_{t^{-1}}(v)(-1)^{Q_0(v)} {\rm Tr}(U_{st} P(t^{-1}v) P(v))
\notag
\\
&=
\eta_{t^{-1}}(v)(-1)^{Q_0(v)+B_0(t^{-1}v, v)} 
\notag
\\
&\quad\times
{\rm Tr}(U_r P((t^{-1}+1)v)),
\label{eq:K_v}
\end{align}
where we have put
\begin{align}
r
&=
st.
\end{align}

Let us first consider the case of $r\neq 1$ (the case of $r=1$ will be treated separately).
We can show that $K_v(s,t)$ is proportional to ${\rm Tr}(U_r)$.
To see this, we evaluate ${\rm Tr}(U_r)$ in the same way
as in Eq.~\eqref{eq:K_v}:
\begin{align}
{\rm Tr}(U_r)
&=
{\rm Tr}(P(v) U_r P(v)^\dagger)
\notag
\\
&=
\eta_{r^{-1}}(v) (-1)^{Q_0(v)}
{\rm Tr}(U_r P(r^{-1}v) P(v))
\notag
\\
&=
\eta_{r^{-1}}(v) (-1)^{Q_0(v)+B_0(r^{-1}v, v)}
\notag
\\
&\quad\times
{\rm Tr}(U_r P((r^{-1}+1)v)).
\label{eq:Tr_U_r}
\end{align}
Denoting the argument of the Pauli string in Eq.~\eqref{eq:Tr_U_r} by $w=(r^{-1}+1)v \in E$,
we obtain
\begin{align}
{\rm Tr}(U_r P(w))
&=
\eta_{r^{-1}}\left(\frac{w}{r^{-1}+1}\right)^{-1}
\notag
\\
&\quad\times
(-1)^{Q_0\left(\frac{w}{r^{-1}+1}\right)+B_0\left(\frac{r^{-1}w}{r^{-1}+1}, \frac{w}{r^{-1}+1}\right)}
{\rm Tr}(U_r),
\end{align}
where we used $r^{-1}+1\neq 0$ (since $r\neq 1$).
By substituting the above relation to Eq.~\eqref{eq:K_v}, we arrive at
\begin{align}
K_v(s,t)
&=
\eta_{t^{-1}}(v)(-1)^{Q_0(v)+B_0(t^{-1}v, v)} 
\notag
\\
&\quad\times
\eta_{r^{-1}}(u)^{-1}(-1)^{Q_0(u)+B_0(r^{-1}u, u)}
{\rm Tr}(U_r),
\label{eq:K_v_2}
\end{align}
where we have put
\begin{align}
u
&=
\frac{t^{-1}+1}{r^{-1}+1}v.
\end{align}
Using the relation \eqref{eq:Tr_U_s} with $r\neq 1$,
we can write Eq.~\eqref{eq:K_v_2} in a compact form,
\begin{align}
K_v(s,t)
&=
-\frac{\eta_{t^{-1}}(v)}{\eta_{r^{-1}}(u)}
(-1)^{E_0(v,u)},
\label{eq:K_v_3}
\end{align}
where we have defined the phase exponent $E_0(v,u)$ as
\begin{align}
E_0(v, u)
&=
Q_0(v)+Q_0(u)+B_0(t^{-1}v, v)+B_0(r^{-1}u, u).
\end{align}
From Eq.~\eqref{eq:K_v_3}, we can see that $K_v(s,t)\in\{\pm 1, \pm i\}$.

We further use the fact that $K_v(s,t)$ is invariant along the orbit in the present Singer-cycle circuit:
\begin{align}
K_v(s,t)
&=
K_{\zeta v}(s,t).
\end{align}
One can quickly confirm this by
\begin{align}
K_v(s,t)
&=
{\rm Tr}(U_s (UP(v)U^\dagger) U_t (UP(v)^\dagger U^\dagger))
\notag
\\
&=
\eta(v) \overline{\eta(v)}
{\rm Tr}(U_s P(\zeta v) U_t P(\zeta v)^\dagger)
\notag
\\
&=
K_{\zeta v}(s,t).
\end{align}
Thus, we can multiply $K_v(s,t)$ over the whole orbit to obtain
\begin{align}
[K_v(s,t)]^N
&=
\prod_{j=0}^{N-1} K_{\zeta^j v}(s,t)
\notag
\\
&=
-\frac{\prod_{j=0}^{N-1} \eta_{t^{-1}}(\zeta^j v)}{\prod_{j=0}^{N-1} \eta_{r^{-1}}(\zeta^j u)}
(-1)^{\sum_{j=0}^{N-1}E_0(\zeta^j v, \zeta^j u)}.
\label{eq:K_v^N}
\end{align}
Now the product of the phase factor $\eta_t$ (with arbitrary $t=\zeta^n$) over the orbit can be evaluated as
\begin{align}
\prod_{j=0}^{N-1} \eta_t(\zeta^j v)
&=
\prod_{j=0}^{N-1} \prod_{k=0}^{n-1}
\eta(\zeta^{j+k} v)
=
\prod_{k=0}^{n-1}\prod_{j=0}^{N-1} 
\eta(\zeta^{j+k} v)
\notag
\\
&=
\left[\prod_{j=0}^{N-1} 
\eta(\zeta^j v)
\right]^n
=
[\eta_{\zeta^N}(v)]^n
=
1.
\end{align}
In the last line above, we used the relation $\eta_{\zeta^N}(v)=1$ (see Sec.~\ref{sec:eigenvalues_U}).
In this way, we can remove the $\eta$ factor from Eq.~\eqref{eq:K_v^N} and obtain
\begin{align}
[K_v(s,t)]^N
&=
-(-1)^{E(v,u)},
\label{eq:K_v^N_2}
\end{align}
where we have defined the sum of the phase exponent $E_0$ over the orbit,
\begin{align}
E(v,u)
&=
\sum_{j=0}^{N-1}E_0(\zeta^j v, \zeta^j u).
\end{align}
Using the sum of $B_0$ and $Q_0$ over the orbit [Eqs.~\eqref{eq:B_def} and \eqref{eq:Q_def}], 
we can write $E(v,u)$ as
\begin{align}
E(v,u)
&=
Q(v)+Q(u)+B(t^{-1}v,v)+B(r^{-1}u,u).
\end{align}
From Eq.~\eqref{eq:K_v^N_2} together with the fact that $N$ is odd,
we find that $K_v(s,t)\in\{\pm 1\}$. In particular, we obtain
\begin{align}
K_v(s,t)
&=
-(-1)^{E(v,u)}.
\end{align}

Next, we rewrite $E(v,u)$ in terms of the finite-field traces.
Let us define
\begin{align}
b
&=
\frac{t^{-1}+1}{r+1},
\end{align}
which leads to $u=brv$.
Using Eqs.~\eqref{eq:B_trace} and \eqref{eq:Q_trace},
we can evaluate $E(v,u)$ as
\begin{align}
E(v,u)
&=
{\rm tr}_E(\beta v \bar v+\beta bv\overline{bv}+\beta t^{-1}v\bar v
+\beta r^{-1}bv \overline{bv})
\notag
\\
&=
{\rm tr}_E(\beta v \bar v(1+b\bar b+t^{-1}+r^{-1}b\bar b)),
\end{align}
where we have used the relation $r\bar r=1$ (since $r\in T$).
By explicitly calculating the argument of the trace as
\begin{align}
1+b\bar b+t^{-1}+r^{-1}b\bar b
&=
1+t^{-1}+\frac{t+t^{-1}}{r+1}
=
1+\frac{s+t}{st+1},
\end{align}
we find that
\begin{align}
E(v,u)
&=
{\rm tr}_E\left(\beta v\bar v\left(1+\frac{s+t}{st+1}\right)\right)
\notag
\\
&=
{\rm tr}_K\left(v\bar v\left(1+\frac{s+t}{st+1}\right)\right).
\end{align}
The last equality follows from
\begin{align}
\overline{\left(1+\frac{s+t}{st+1}\right)}
&=
1+\frac{s+t}{st+1}
\in
K
\end{align}
and $\beta+\bar\beta=1$ [Eq.~\eqref{eq:beta}].

Let us define the additive character $\psi: K\to \{\pm 1\}$ by
\begin{align}
\psi(x)
&=
(-1)^{{\rm tr}_K(x)}
\quad
(x\in K).
\end{align}
The additivity of the trace implies
\begin{align}
\psi(x+y)
&=
\psi(x)\psi(y),
\end{align}
so $\psi$ is a character of the additive group $(K,+)$.
In terms of $\psi$, the kernel $K_v(s,t)$ takes the form
\begin{align}
K_v(s,t)
&=
-\psi\left(v\bar v\left(1+\frac{s+t}{st+1}\right)\right)
\quad
(st\neq 1).
\end{align}

Finally, let us consider the case of $r=st=1$.
For $s=t=1$, one can immediately see that
\begin{align}
K_v(s,t)
&=
{\rm Tr}(P(v)P(v)^\dagger)
=
D.
\end{align}
For $st=1$ and $t\neq 1$, we notice from Eq.~\eqref{eq:K_v}
that $K_v(s,t)$ is proportional to
\begin{align}
{\rm Tr}(U_r P((t^{-1}+1)v))
&=
{\rm Tr}(P((t^{-1}+1)v)),
\end{align}
which vanishes since $(t^{-1}+1)v\neq 0$.
To summarize, we have obtained a closed form of $K_v(s,t)$:
\begin{align}
K_v(s,t)
&=
\begin{cases}
D & (s=t=1) \\
0 & (st=1, t\neq 1) \\
\displaystyle
-\psi\left(v\bar v\left(1+\frac{s+t}{st+1}\right)\right) & (st\neq 1)
\end{cases}.
\end{align}
Using this result, we can write the squared matrix elements of Pauli strings
as
\begin{align}
|\langle \chi| P(v)|\chi'\rangle|^2
&=
\frac{D-X_{\chi,\chi'}(v)}{N^2},
\end{align}
where we define
\begin{align}
X_{\chi,\chi'}(v)
&=
\sum_{\substack{s,t\in T \\ st\neq 1}}
\overline{\chi(s)}\, \overline{\chi'(t)}
\psi\left(v\bar v\left(1+\frac{s+t}{st+1}\right)\right).
\label{eq:X_def}
\end{align}

\subsection{Transformation from the double sum to a single sum}

So far, the finite-field calculation is straightforward and elementary.
A crucial step now is to transform the double sum that appeared in Eq.~\eqref{eq:X_def}
into a single sum over the finite field $E$. 
This allows us to evaluate the squared matrix elements
with Weil bounds on the sum of mixed characters \cite{Weil1948, Castro2000, Cochrane2006}, which is the only non-elementary
mathematical input that we will use for the derivation of the strong ETH \eqref{eq:infinite_T_ETH}
(we will use a similar Weil bound for additive character sums in Sec.~\ref{sec:temporal_fluctuation}).

To see this, let us consider the following domain $\mathcal D\subset T\times T$,
\begin{align}
\mathcal D
&=
\{(s,t)\in T\times T \, | \, s\neq 1, t\neq 1, st\neq 1\}.
\end{align}
We define a map $F: \mathcal D\to E\backslash K$ with
\begin{align}
F:
(s,t)
\mapsto
x
=
\frac{s+1}{st+1}.
\label{eq:F_def}
\end{align}
One can confirm that the image of the map $F$ belongs to $E\backslash K$ as follows.
The conjugate of $x$ reads
\begin{align}
\bar x
&=
\overline{\left(\frac{s+1}{st+1}\right)}
=
\frac{\bar s+1}{\overline{st}+1}
=
\frac{s^{-1}+1}{s^{-1}t^{-1}+1}
\notag
\\
&=
\frac{t(s+1)}{st+1}
=
tx.
\label{eq:bar_x}
\end{align}
Since $x\neq 0$ (as follows from $s\neq 1$), we find that $\bar x\neq x$,
which implies that $x\notin K$.

We also find that
\begin{align}
x+1
&=
\frac{s(t+1)}{st+1}
=
s(\bar x+1).
\label{eq:x+1}
\end{align}
Since $x\neq 0, 1$, we can take the inverse of $F$ from Eqs.~\eqref{eq:bar_x} and \eqref{eq:x+1},
\begin{align}
F^{-1}:
x
&\mapsto
(s,t)
=
\left(\frac{x+1}{\bar x+1}, \frac{\bar x}{x}\right).
\end{align}
In this way, $F$ gives a one-to-one mapping between $\mathcal D$ and $E\backslash K$.

In order to use this map, we rewrite the double sum in Eq.~\eqref{eq:X_def} as
\begin{align}
X_{\chi,\chi'}(v)
&=
\sum_{(s,t)\in \mathcal D}
\overline{\chi(s)}\, \overline{\chi'(t)}
\psi\left(v\bar v\left(1+\frac{s+t}{st+1}\right)\right)
\notag
\\
&\quad
+\left(\sum_{\substack{s=1 \\ t\neq 1}}
+\sum_{\substack{s\neq 1 \\ t=1}}\right)
\overline{\chi(s)}\, \overline{\chi'(t)}
\psi\left(v\bar v\left(1+\frac{s+t}{st+1}\right)\right).
\label{eq:X_v}
\end{align}
For $s=1$ and $t\neq 1$, the argument of $\psi$ vanishes ($1+\frac{s+t}{st+1}=0$),
and we obtain
\begin{align}
&
\sum_{\substack{s=1 \\ t\neq 1}}
\overline{\chi(s)}\, \overline{\chi'(t)}
\psi\left(v\bar v\left(1+\frac{s+t}{st+1}\right)\right)
\notag
\\
&=
\sum_{t\neq 1} \overline{\chi'(t)}
=
\sum_{t\in T} \overline{\chi'(t)}-1
=
-1.
\end{align}
The last equality follows from the orthogonality of the character $\chi$ [Eq.~\eqref{eq:orthogonality_chi}]
and the fact that the trivial character does not appear as a label of the eigenstates in the Singer-cycle circuit
(see the discussion in Sec.~\ref{sec:eigenvalues_U}).
A similar evaluation applies to the sum over $s\neq 1$ and $t=1$ in Eq.~\eqref{eq:X_v}.
Hence we arrive at the expression,
\begin{align}
X_{\chi,\chi'}(v)
&=
\sum_{(s,t)\in \mathcal D}
\overline{\chi(s)}\, \overline{\chi'(t)}
\psi\left(v\bar v\left(1+\frac{s+t}{st+1}\right)\right)
-2.
\label{eq:X_v_2}
\end{align}

Now we use the map $F$ \eqref{eq:F_def} to rewrite Eq.~\eqref{eq:X_v_2}
in terms of the single sum over $E\backslash K$. By noting that
\begin{align}
x+\bar x
&=
1+\frac{s+t}{st+1},
\end{align}
we can write the additive-character part as
\begin{align}
\psi\left(v\bar v\left(1+\frac{s+t}{st+1}\right)\right)
&=
\psi(v\bar v (x+\bar x))
=
(-1)^{{\rm tr}_K(v\bar v(x+\bar x))}
\notag
\\
&=
(-1)^{{\rm tr}_E(v\bar vx)}.
\end{align}
Let us define the additive character of $(E,+)$ by
\begin{align}
\Psi(u)
&=
(-1)^{{\rm tr}_E(u)}
\quad
(u\in E),
\label{eq:Psi_def}
\end{align}
which satisfies
\begin{align}
\Psi(u+v)
&=
\Psi(u)\Psi(v)
\quad
(u, v\in E).
\end{align}
With the character $\Psi$, $X_{\chi,\chi'}(v)$ \eqref{eq:X_v_2} can be expressed as
\begin{align}
X_{\chi,\chi'}(v)
&=
\sum_{(s,t)\in \mathcal D}
\overline{\chi(s)}\, \overline{\chi'(t)}
\Psi(v\bar v x)-2.
\label{eq:X_v_3}
\end{align}

The remaining step is to replace the characters $\chi$ and $\chi'$ of $T$
by the character of the cyclic group $E^\times=\{1, g, g^2, \dots, g^{D^2-2}\}$ 
($g$ is a generator of $E^\times$) defined by
\begin{align}
\rho(g^n)
&=
e^{2\pi in/(D^2-1)}
\quad
(n=0,1,\dots,D^2-2).
\label{eq:rho_def}
\end{align}
The character $\rho$ satisfies $\rho(uv)=\rho(u)\rho(v)$ ($u, v\in E^\times$).
As shown in Sec.~\ref{sec:gate_coefficients}, one can take $g$
in such a way that $g^{D-1}=\zeta\in T$.
If we write $\chi=\chi_k$ ($k=0,1,\dots,N-1$),
the character $\chi_k$ is related to $\rho$ \eqref{eq:rho_def} through
\begin{align}
\chi_{k}(\zeta^n)
&=
e^{2\pi i kn/(D+1)}
=
e^{2\pi ikn(D-1)/(D^2-1)}
\notag
\\
&=
\rho(g^{kn(D-1)})
=
\rho(\zeta^{kn})
=
\rho(\zeta^n)^k.
\label{eq:chi_rho}
\end{align}

The variables $s, t\in T$ can be expressed in terms of $x\in E\backslash K$ as
\begin{align}
s
&=
\frac{x+1}{\bar x+1}
=
\frac{x+1}{x^D+1}
=
\frac{x+1}{(x+1)^D}
=
(x+1)^{-D+1}
\end{align}
and
\begin{align}
t
&=
\frac{\bar x}{x}
=
x^{D-1}.
\end{align}
Using these expressions, we can write the characters in Eq.~\eqref{eq:X_v_3} 
(with $\chi=\chi_k$ and $\chi'=\chi_{k'}$) as
\begin{align}
\overline{\chi_k(s)}\, \overline{\chi_{k'}(t)}
&=
\chi_k(s^{-1})\overline{\chi_{k'}(t)}
\notag
\\
&=
\chi_k((x+1)^{D-1})
\overline{\chi_{k'}(x^{D-1})}.
\end{align}
We further replace $\chi_k$ and $\chi_{k'}$ with $\rho$ 
through Eq.~\eqref{eq:chi_rho}, 
\begin{align}
\overline{\chi_k(s)}\, \overline{\chi_{k'}(t)}
&=
\rho((x+1)^{D-1})^k
\overline{\rho(x^{D-1})^{k'}}
\notag
\\
&=
\rho(x+1)^{k(D-1)}
\rho(x)^{-k'(D-1)}
\end{align}
If we define the order of $E^\times$ by
\begin{align}
m
&=
D^2-1,
\end{align}
then $x^m=1$ for $x\in E\backslash K$.
Therefore, the characters in Eq.~\eqref{eq:X_v_3} are written
with respect to $\rho$ of a polynomial of $x$,
\begin{align}
\overline{\chi_k(s)}\, \overline{\chi_{k'}(t)}
&=
\rho((x+1)^{k(D-1)}x^{m-k'(D-1)}).
\end{align}

With the one-to-one mapping $F$ \eqref{eq:F_def},
the sum over the domain $\mathcal D$ in Eq.~\eqref{eq:X_v_3} is replaced by
the sum over $E\backslash K$,
\begin{align}
X_{\chi,\chi'}(v)
&=
\sum_{x\in E\backslash K}
\rho((x+1)^{k(D-1)}x^{m-k'(D-1)})\Psi(v\bar vx)
\notag
\\
&\quad
-2.
\end{align}
Finally, we change the sum over $E\backslash K$ to the one over $E\backslash \{0,1\}$.
The difference amounts to a sum over $K\backslash\{0,1\}$, in which
$x=\bar x$,
$\Psi(v\bar vx)=(-1)^{{\rm tr}_K(v\bar v(x+\bar x))}=1$,
$x^{D-1}=1$, and $(x+1)^{D-1}=(\bar x+1)/(x+1)=1$.
Therefore, we obtain
\begin{align}
\sum_{x\in K\backslash \{0,1\}}
\rho((x+1)^{k(D-1)}x^{m-k'(D-1)})\Psi(v\bar vx)
&=
D-2,
\end{align}
which leads to
\begin{align}
X_{\chi,\chi'}(v)
&=
\sum_{x\in E\backslash \{0,1\}}
\rho((x+1)^{k(D-1)}x^{m-k'(D-1)})\Psi(v\bar vx)
\notag
\\
&\quad
-D.
\end{align}

As a result, we arrive at the final expression for the squared matrix elements
of non-identity Pauli strings on the eigenstate basis:
\begin{align}
|\langle \psi_k| P(v) |\psi_{k'}\rangle|^2
&=
\frac{2D-Y_{k,k'}(v)}{N^2}.
\end{align}
Here $k,k'=1,2,\dots, D$ (we exclude $k,k'=0$ because an eigenstate of $U$ with the eigenvalue 1 is absent) and the function $Y_{k,k'}(v)$ is defined in terms of a sum of mixed characters
(i.e., a product of the multiplicative character $\rho$ and the additive character $\Psi$),
\begin{align}
Y_{k,k'}(v)
&=
\sum_{x\in E\backslash \{0,1\}}
\rho((x+1)^{k(D-1)}x^{m-k'(D-1)})\Psi(v\bar vx).
\label{eq:Y_{k,k'}}
\end{align}
This form is suitable for the application of Weil bounds on a sum of mixed characters,
as we will see in the next subsection.

\subsection{Estimate on the sum of mixed characters and strong ETH}
\label{sec:mixed_sum}

Now we use the following Weil bound for mixed character sums \cite{Weil1948, Castro2000, Cochrane2006},
\begin{align}
\left|
{\sum_{x\in\mathbb F_{2^n}}}'
\rho(f(x))\Psi(g(x))
\right|
&\le
(s+d-1)2^{n/2}.
\label{eq:Weil_bound}
\end{align}
Here $\rho$ and $\Psi$ are nontrivial multiplicative and additive characters of $\mathbb F_{2^n}$, respectively, $f$ and $g$ are nonzero polynomials over $\mathbb F_{2^n}$, and the sum is taken over $\mathbb F_{2^n}$ except for zeros of $f(x)$. 
We assume that $\Psi(g(x))$ is nonconstant
on $\mathbb F_{2^n}$ and that the multiplicity of each root of $f$ is not divisible by the order of $\rho$.
On the right side of Eq.~\eqref{eq:Weil_bound},
$s$ is the number of distinct roots of $f(x)$ in the algebraic closure of $\mathbb F_{2^n}$,
and $d (\ge 1)$ is the degree of the polynomial $g(x)$.

We apply this bound to $Y_{k,k'}(v)$ \eqref{eq:Y_{k,k'}} 
over the field $E=\mathbb F_{D^2}$ by taking
$f(x)
=
(x+1)^{k(D-1)}x^{m-k'(D-1)}$ and
$g(x)=v\bar v x$,
where $m=D^2-1$ is the order of $\rho$.
For $1\le k,k'\le D$, the polynomial $f$ has two distinct roots, $0$ and $1$, with multiplicities
$m-k'(D-1)$ $(<m)$ and $k(D-1)$ $(<m)$,
which are not divisible by the order of $\rho$.
Since $\Psi$ is a nontrivial additive character
of $E$, the function $x\mapsto\Psi(v\bar v x)$ is
nonconstant for $v\neq 0$.
We can therefore apply the Weil bound \eqref{eq:Weil_bound} with $s=2$ and $d=1$ to our case, giving
\begin{align}
|Y_{k,k'}(v)|
&\le
2D
\quad
(v\neq 0)
\end{align}
for all $k,k'=1,2,\dots,D$. 

As a consequence, the matrix elements of non-identity Pauli operators are bounded uniformly for all the eigenstates as
\begin{align}
|\langle \psi_k| P(v) |\psi_{k'}\rangle|^2
&\le
\frac{4D}{N^2},
\end{align}
or equivalently
\begin{align}
|\langle \psi_k| P(v) |\psi_{k'}\rangle|
&\le
\frac{2\sqrt{D}}{D+1}.
\label{eq:Pauli_bound}
\end{align}
This is the main result of the paper, serving as a counterpart of the bound \eqref{eq:Pauli_eval2} on the matrix elements of Pauli operators in the LFSR circuit. The crucial difference is that the bound \eqref{eq:Pauli_bound} in our Singer-cycle circuit is valid for all the eigenstates ($k,k'=1,2,\dots,D$),
while the bound \eqref{eq:Pauli_eval2} in the LFSR circuit holds for almost all eigenstates except for two degenerate eigenstates ($k,k'=1,2,\dots,D-2$).

The derivation from the bound \eqref{eq:Pauli_bound} to the strong ETH \eqref{eq:infinite_T_ETH} follows the same argument as in Ref.~\cite{IppolitiLong2026}, which we briefly summarize here.
Any observable $O$ supported on $\ell$ sites (which are not necessarily distributed locally) can be
expanded in terms of Pauli strings $P$ on those sites,
\begin{align}
O
&=
\sum_P c_P P.
\label{eq:O_expansion}
\end{align}
The sum includes the identity operator $I$
with the coefficient $c_I=\operatorname{Tr}O/D$.
The triangle inequality then gives
\begin{align}
\left|
\langle\psi_k|O|\psi_{k'}\rangle
-\frac{\operatorname{Tr}O}{D}\delta_{k,k'}
\right|
&=
\left|
\sum_{P\neq I} c_P \langle \psi_k|P|\psi_{k'}\rangle
\right|
\notag
\\
&\le
\sum_{P\ne I}|c_P|
|\langle \psi_k|P|\psi_{k'}\rangle|
\notag
\\
&\le
\frac{2\sqrt{D}}{D+1} \sum_{P\neq I} |c_P|.
\end{align}
The last inequality follows from the bound \eqref{eq:Pauli_bound}.

From the Cauchy--Schwarz inequality, we have
\begin{align}
\left(
\sum_{P\neq I} |c_P|
\right)^2
\le
\sum_{P\neq I} |c_P|^2 \sum_{P\neq I} 1^2
=
(4^\ell-1)\sum_{P\neq I} |c_P|^2.
\end{align}
The orthogonality of the Pauli basis,
${\rm Tr}(P^\dagger P')=D\delta_{P,P'}$, implies
\begin{align}
\sum_{P\ne I}|c_P|^2
&=
\frac{1}{D}\operatorname{Tr}(O^\dagger O)
-\left|\frac{\operatorname{Tr}O}{D}\right|^2
\le
\|O\|_{\mathrm{op}}^2,
\label{eq:c_P^2}
\end{align}
where $\|O\|_{\mathrm{op}}$ denotes the operator norm.

By combining these inequalities, we obtain an estimate 
\begin{align}
\left|
\langle\psi_k|O|\psi_{k'}\rangle
-\frac{\operatorname{Tr}O}{D}\delta_{k,k'}
\right|
\le
\frac{2\sqrt{D}}{D+1}
\sqrt{4^\ell-1}\|O\|_{\mathrm{op}}.
\label{eq:strong_ETH}
\end{align}
For fixed $\ell$ and bounded operator $O$, the right-hand side vanishes as $O(D^{-1/2})$ in the thermodynamic limit uniformly in $k$ and $k'$.
Thus, all the eigenstates reproduce the infinite-temperature expectation values of arbitrary finite-support observables,
establishing the strong ETH \eqref{eq:infinite_T_ETH}.
The off-diagonal matrix elements vanish uniformly as well.

Finally, let us comment that there are no nontrivial local conserved quantities in the present Singer-cycle circuit. This can be understood in the same way as in the LFSR circuit (see Sec.~\ref{sec:LFSR_circuit}).
Suppose that there is a local conserved quantity $Q$.
We expand it in terms of Pauli strings,
$Q=\sum_{\bm v} c_{\bm v} P(\bm v)$.
From Eq.~\eqref{eq:clifford_action} together with $UQU^\dagger=Q$, we find that $c_{M\bm v}=\eta(\bm v)c_{\bm v}$. Thus, the set of nonzero coefficients must contain the entire orbit of Pauli strings.
In the Singer-cycle circuit, the length of each orbit is $D+1$. On the other hand, if each local Pauli operator $P(\bm v)$ acts on at most $\ell$ adjacent sites, the number of terms appearing in $Q$ is at most $L(4^\ell-1)$.
Since $L(4^\ell-1)<D+1$ for sufficiently large systems, we conclude that there is no such kind of $Q$.

\section{Temporal fluctuation of expectation values of observables}
\label{sec:temporal_fluctuation}

In the previous section, we have established the strong version of ETH \eqref{eq:infinite_T_ETH} in the Singer-cycle circuits. This ensures that the long-time average of an expectation value of any observable with a finite support converges to the thermal value at infinite temperature. However, it does not tell whether the expectation value really stays close to the thermal value 
for most times, since the temporal fluctuation around the thermal value may not be suppressed. 

The question is especially nontrivial in the case of Clifford circuits, since if the initial state is a Stabilizer state 
the magnitude of the expectation value of any Pauli operators is either 
$|\langle P(v)\rangle_t|=0$ or 1 at every time step. On the other hand, the thermal value of (non-identity) Pauli operators at infinite temperature vanishes. Therefore, if the expectation value approaches a thermal value at most of the times, then the time interval at which $|\langle P(v)\rangle_t|=1$ should be negligible. We also remark that the Singer-cycle circuit shows an exact recurrence due to $U^N=I$ (the recurrence time is $N=2^L+1$), so that expectation values cannot stay at thermal values during the whole time steps.

In this section, we prove that the time average of temporal fluctuation of expectation values of observables vanishes in the large system-size limit.
More precisely, we will show that
\begin{align}
\overline{|\langle O\rangle_t-O_{\rm th}|^2}
&\le
O(D^{-1/2}),
\end{align}
where $\langle O\rangle_t$ denotes the expectation value of an observable $O$ at time step $t\in \mathbb Z$, and the overlines represent the time average. Thus, the root-mean-square deviation from the thermal value is bounded by $O(D^{-1/4})$.

Let us first consider the case that the initial state is given by a pure state, $\rho=|\psi\rangle \langle\psi|$. We expand the density matrix in the Pauli-operator basis,
\begin{align}
\rho
&=
\frac{1}{D}\sum_v r_v^\ast P(v),
\label{eq:rho_expansion}
\end{align}
where the expansion coefficient can be written as
\begin{align}
r_v
&=
{\rm Tr}[\rho P(v)],
\end{align}
and we recall that the Pauli-operator basis satisfies the orthogonality relation, ${\rm Tr}[P(v)^\dagger P(w)]=D\delta_{v,w}$. Since ${\rm Tr}(\rho^2)=1$, we have a normalization condition, $\sum_v |r_v|^2=D$.

The expectation value of a Pauli operator $P(v)$ ($v\neq 0$) at time step $t\in\mathbb Z$ is given by
\begin{align}
\langle P(v)\rangle_t
&=
{\rm Tr}[\rho U^{-t}P(v)U^t].
\end{align}
Then we evaluate the time average of the temporal fluctuation of it, which is defined by
\begin{align}
\overline{|\langle P(v)\rangle_t|^2}
&=
\frac{1}{N}\sum_{t=0}^{N-1} 
\left|
{\rm Tr}[\rho U^{-t}P(v)U^t]
\right|^2
\notag
\\
&=
\frac{1}{N}\sum_{s\in T}
\left|
{\rm Tr}[\rho P(sv)]
\right|^2.
\label{eq:time_average_P^2}
\end{align}
In the second line in the above, we canceled the phase factor $\eta$ appearing in front of the Pauli operator after $t$ steps evolution, and replaced the sum over $t$ with the sum over the cyclic group $T=\{1,\zeta, \dots, \zeta^{N-1}\}$ \eqref{eq:group_T}.
After substituting the expansion \eqref{eq:rho_expansion} in Eq.~\eqref{eq:time_average_P^2}, we obtain
\begin{align}
\overline{|\langle P(v)\rangle_t|^2}
&=
\frac{1}{N}\sum_{s\in T} |r_{sv}|^2.
\label{eq:time_average_P^2_2}
\end{align}

Now we rewrite $|r_{sv}|^2$ in the following way.
Since the initial state is pure, we have
\begin{align}
|r_v|^2
&=
{\rm Tr}[\rho P(v)] {\rm Tr}[\rho P(v)^\dagger]
=
\langle \psi|P(v)|\psi\rangle
\langle \psi|P(v)^\dagger|\psi\rangle
\notag
\\
&=
{\rm Tr}[\rho P(v)\rho P(v)^\dagger]
\notag
\\
&=
\frac{1}{D^2}\sum_{w,w'} r_w^\ast r_{w'}
{\rm Tr}[P(w) P(v) P(w')^\dagger P(v)^\dagger].
\end{align}
By using the relations \eqref{eq:pauli_commutation} and \eqref{eq:P_dagger}, we can transform the above expression into
\begin{align}
|r_v|^2
&=
\frac{1}{D}\sum_w |r_w|^2 (-1)^{J(v,w)}.
\label{eq:c_v^2}
\end{align}
The idea behind this rewriting is that there is a cancellation in the summation of the phase factor $(-1)^{J(sv,w)}$, which can be estimated by the Weil bound. Substituting the relation \eqref{eq:c_v^2} in Eq.~\eqref{eq:time_average_P^2_2}, we obtain
\begin{align}
\overline{|\langle P(v)\rangle_t|^2}
&=
\frac{1}{N}\sum_{s\in T} 
\frac{1}{D}\sum_w |r_w|^2 (-1)^{J(sv,w)}
\notag
\\
&=
\frac{1}{D}+\frac{1}{ND}
\sum_{w\neq 0} |r_w|^2 \sum_{s\in T} (-1)^{J(sv,w)}.
\label{eq:time_average_P^2_3}
\end{align}
The first term in the second line comes from the contribution of $w=0$ (note that $r_0=1$ and $J(sv,0)=0$).

In our phase convention, the symplectic form is written in terms of the trace over the field $E$ as
$J(sv,w)={\rm tr}_E(sv\overline{w})$ (see Eq.~\eqref{eq:J_trace}). Hence the phase factor in Eq.~\eqref{eq:time_average_P^2_3} becomes the additive character of the field $E$ \eqref{eq:Psi_def}, $(-1)^{J(sv,w)}=\Psi(sv\overline{w})$.
Furthermore, the sum of the character $\Psi$ over $T$ can be replaced by the Kloosterman sum of the character $\psi$ of the field $K$ \cite{Kloosterman1928},
\begin{align}
\sum_{s\in T}\Psi(as)
&=
-\sum_{x\in K^\times} \psi\left(x+\frac{a\bar a}{x}\right)
\quad
(a\in E, a\neq 0).
\label{eq:Psi_sum}
\end{align}

To show this relation, let us put $u=as\in E$, which visits each element in $E$ that satisfies $u\bar u=a\bar a$ once (note that $s\bar s=1$).
If we denote the left hand side of Eq.~\eqref{eq:Psi_sum} by $S(a)=\sum_{s\in T} \Psi(as)$, then
\begin{align}
S(a)
&=
\sum_{\substack{u\in E \\ u\bar u=a\bar a}}
\Psi(u)
=
\sum_{\substack{u\in E \\ u\bar u=a\bar a}}
\psi(u+\bar u).
\end{align}
If we define the number $N_1(a,b)$ ($a\in E, b\in K$) as
\begin{align}
N_1(a,b)
&=
\#\{ u\in E :
u\bar u=a\bar a, u+\bar u=b\},
\end{align}
then $S(a)$ takes the form of the sum over the field $K$,
\begin{align}
S(a)
&=
\sum_{b\in K} N_1(a,b)\psi(b).
\end{align}

Now we compare $N_1(a,b)$ with another number $N_2(a,b)$ defined by
\begin{align}
N_2(a,b)
&=
\#\{x\in K^\times: x+a\bar a/x=b\}.
\end{align}
We find that the equality
\begin{align}
N_1(a,b)+N_2(a,b)
&=
2
\end{align}
holds for arbitrary $a\neq 0$ and $b$.
In fact, the two numbers, $N_1$ and $N_2$,
are related to roots of the polynomial $t^2+bt+a\bar a$.
For $b\ne0$, the polynomial either has two roots in $K$ or has two conjugate roots in $E\setminus K$. In the former case, we have $N_1=0$ and $N_2=2$, while in the latter $N_1=2$ and $N_2=0$.
For $b=0$, the polynomial has the unique solution $\sqrt{a\bar a}\in K^\times$ and hence $N_1=N_2=1$.
As a result, we arrive at
\begin{align}
S(a)
&=
\sum_{b\in K} (2-N_2(a,b))\psi(b)
=
-\sum_{x\in K^\times} \psi\left(x+\frac{a\bar a}{x}\right),
\end{align}
where we used the orthogonality of the additive character $\psi$. Thus, the relation \eqref{eq:Psi_sum} is proved.

Therefore, the phase part of Eq.~\eqref{eq:time_average_P^2_3} is written in terms of the Kloosterman sum as
\begin{align}
\overline{|\langle P(v)\rangle_t|^2}
&=
\frac{1}{D}-\frac{1}{ND}
\sum_{w\neq 0} |r_w|^2 \sum_{x\in K^\times} 
\psi\left(x+\frac{v\bar v w\bar w}{x}\right).
\label{eq:time_average_P^2_4}
\end{align}
Here we use the Weil bound for the Kloosterman sum \cite{Castro2000, Cochrane2006, Conrad2002},
\begin{align}
\left|
\sum_{x\in K^\times}
\psi\left(x+\frac{n}{x}\right)
\right|
\le
2\sqrt{D},
\end{align}
with $n\in K^\times$. This is a special version of the Weil bound for general additive character sums.
Using the Weil bound, we obtain an estimate
\begin{align}
\overline{|\langle P(v)\rangle_t|^2}
&\le
\frac{1}{D}+\frac{2\sqrt{D}}{ND}\sum_{w\neq 0} |r_w|^2
=
\frac{1}{D}+\frac{2(D-1)}{(D+1)\sqrt{D}},
\label{eq:fluctuation_bound}
\end{align}
which vanishes in the thermodynamic limit as $O(D^{-1/2})$.
This result indicates that the expectation value of Pauli operators is arbitrarily close to the thermal value at almost every time step. 

So far, we have focused on the situation that the initial state is a pure state. We can generalize our argument to general mixed states $\rho=\sum_\alpha p_\alpha |\psi_\alpha\rangle \langle \psi_\alpha|$ by using the convexity of squared expectation values,
\begin{align}
\left|
\sum_\alpha p_\alpha \langle P(v)\rangle_{\rho_\alpha,t}
\right|^2
\le
\sum_\alpha p_\alpha 
\left|
\langle P(v)\rangle_{\rho_\alpha,t}
\right|^2,
\end{align}
where $\langle P(v)\rangle_{\rho_\alpha,t}$ denotes the expectation value of $P(v)$ at time step $t$ with the initial pure state $\rho_\alpha=|\psi_\alpha\rangle \langle\psi_\alpha|$.
Since the fluctuation is uniformly bounded for a pure state as in \eqref{eq:fluctuation_bound}, it is also bounded for mixed states,
\begin{align}
\overline{\left|
\langle P(v)\rangle_{\rho,t}
\right|^2}
\le
B_D,
\label{eq:P_fluctuation}
\end{align}
where $B_D=\frac{1}{D}+\frac{2(D-1)}{(D+1)\sqrt{D}}=O(D^{-1/2})$ denotes the right hand side of Eq.~\eqref{eq:fluctuation_bound}.

Finally, we show that our argument can be applied to general observables supported on a fixed number of sites. Let us consider an observable $O$ that has a support of $\ell$ sites. If we expand $O$ in the Pauli-operator basis as in \eqref{eq:O_expansion}, then the temporal fluctuation of the expectation value around the thermal value $O_{\rm th}=\operatorname{Tr}O/D$ is evaluated as
\begin{align}
\overline{|\langle O\rangle_t-O_{\rm th}|^2}
&=
\overline{\Big|\sum_{P\neq I} c_P \langle P\rangle_t\Big|^2}
\le
\sum_{P\neq I} |c_P|^2
\sum_{P\neq I} \overline{|\langle P\rangle_t|^2}.
\end{align}
In the above, we used the Cauchy-Schwarz inequality. Since $\overline{|\langle P\rangle_t|^2}$ is bounded uniformly as in \eqref{eq:P_fluctuation} and $\sum_{P\neq I} |c_P|^2$ is bounded by the operator norm \eqref{eq:c_P^2}, we find
\begin{align}
\overline{|\langle O\rangle_t-O_{\rm th}|^2}
&\le
(4^\ell-1)\|O\|_{\rm op}^2
B_D,
\label{eq:O_fluctuation}
\end{align}
which vanishes in the thermodynamic limit as $O(D^{-1/2})$.
The inequality \eqref{eq:O_fluctuation} suggests that the number of time steps at which
the expectation value deviates from the thermal value is suppressed by $O(D^{-1/2})$ as
\begin{align}
\frac{\#\{0\le t<n : |\langle O\rangle_t-O_{\rm th}|>\epsilon \}}{n}
&\le
\frac{(4^\ell-1)\|O\|_{\rm op}^2 B_D}{\epsilon^2}.
\end{align}

We also remark that the estimate \eqref{eq:O_fluctuation} leads to the strong ETH with a weaker bound scaling of $O(D^{-1/4})$.
To see this, we apply the Cauchy--Schwarz inequality to the time average, obtaining
\begin{align}
|\overline{\langle O\rangle_t}-O_{\rm th}|^2
&\le
\overline{|\langle O\rangle_t-O_{\rm th}|^2}.
\end{align}
Using the bound \eqref{eq:O_fluctuation}, we find that
\begin{align}
|\overline{\langle O\rangle_t}-O_{\rm th}|
&\le
\sqrt{4^\ell-1} \|O\|_{\rm op}
\sqrt{B_D}
\end{align}
holds for arbitrary initial states.
If we take an eigenstate of $U$ as the initial state, then we recover the strong ETH with the bound proportional to $\sqrt{B_D}=O(D^{-1/4})$.
This provides an alternative (and simpler) proof of the strong ETH for the Singer-cycle circuits but with a weaker estimate than \eqref{eq:strong_ETH}.

A similar bound can be obtained for macroscopic observables as well. Let us write a macroscopic observable as
\begin{align}
A
&=
\sum_{j=1}^L O_j,
\end{align}
where $O_j$ is an observable acting on at most $\ell$ sites. We also assume that $\|O_j\|_{\rm op}\le C$, where $C$ does not depend on $L$. The temporal fluctuation of $\langle O_j\rangle_t$ is bounded as in \eqref{eq:O_fluctuation},
\begin{align}
\overline{|\langle O_j\rangle_t-O_{j,{\rm th}}|^2}
&\le
(4^\ell-1)C^2
B_D.
\end{align}
By using the Cauchy-Schwarz inequality, we find 
\begin{align}
\overline{|\langle A\rangle_t-A_{\rm th}|^2}
&=
\overline{\Big|\sum_{j=1}^L (\langle O_j\rangle_t-O_{j,{\rm th}})\Big|^2}
\notag
\\
&\le
\overline{\sum_{j=1}^L |1|^2 \sum_{j=1}^L |\langle O_j\rangle_t-O_{j,{\rm th}}|^2}
\notag
\\
&\le
L^2(4^\ell-1)C^2
B_D.
\end{align}
Again, the temporal fluctuation of the expectation value of a macroscopic observable is exponentially suppressed in the large system-size limit for arbitrary initial states.

\section{Summary and outlook}
\label{sec:summary}

In this paper, we have constructed a family of brickwork Floquet Clifford circuits that satisfy the strong ETH at infinite temperature \eqref{eq:infinite_T_ETH}. By choosing the circuit coefficients so that the evolution of Pauli operators implements a Singer cycle, we obtain uniform bounds on eigenstate expectation values and off-diagonal matrix elements of observables supported on a fixed number of sites. These bounds scale as $D^{-1/2}$, where $D=2^L$ is the Hilbert-space dimension, and hold for all eigenstates. The proof combines the finite-field structure of squared matrix elements of observables with Weil bounds for mixed character sums, providing an explicit construction of a quantum many-body model in which the strong ETH can be established rigorously for all eigenstates without exceptions.
This is in contrast to the LFSR circuits studied in Ref.~\cite{IppolitiLong2026},
which exhibit weak ergodicity breaking with ETH holding outside a two-dimensional
nonthermal eigenspace.
We have also shown that temporal fluctuations of expectation values of observables vanish in the thermodynamic limit, suggesting that they remain close to the thermal values for almost all times in the thermodynamic limit,
irrespective of the choice of initial states.

There are several directions that remain open. In the Singer-cycle circuits, one has to choose gate coefficients appropriately in a system-size dependent way, which breaks spatial translation symmetry.
While sparse gate configurations with a small number of CNOT gates would suffice even for larger systems,
there is no simple periodic pattern in the coefficients.
Therefore, the question is whether one can construct 
uniform, translation-invariant quantum many-body models that can be shown to satisfy the strong ETH. 
Another direction is to consider systems with nonuniform eigenenergy spacings. 
In the Singer-cycle circuits, the eigenvalues of the Floquet operator distribute on the unit circle on the complex plane with equal spacings (except for the two eigenvalues adjacent to the missing eigenvalue 1). 
While the spectral statistics is commonly associated with quantum chaos, the present circuits show thermal properties of eigenstates. Establishing similarly uniform eigenstate bounds in systems with less constrained spectra will further clarify the relation between thermalization and quantum chaotic aspects of quantum many-body systems.

\acknowledgements

This work is supported by JSPS KAKENHI (Grant Nos.~JP24H00191, JP25H01246, and JP25H01251)
and JST FOREST (Grant No.~JPMJFR2131).
The author used OpenAI chatGPT-6 Astra for the support of mathematical proofs, coding for numerical search, and manuscript revision. The author critically reviewed and independently verified all outputs, and retain full responsibility for this work.

\bibliography{ref}

\begin{thebibliography}{38}
\bibitem [{Cli()}]{Clifford_ordering}%
  \BibitemOpen
  \href@noop {} {}\bibinfo {note} {One can rearrange the shift register
  configuration into a brickwork circuit by cyclicly permuting Clifford gates,
  which can be represented by a unitary transformation.}\BibitemShut {Stop}%
\bibitem [{bou()}]{bound_scaling}%
  \BibitemOpen
  \href@noop {} {}\bibinfo {note} {The scaling of $D^{-1/2}$ is natural since
  \[ \frac{1}{D^2}\sum_{k,k'} \left|\langle\psi_k|O|\psi_{k'}\rangle
  -\frac{\operatorname{Tr}O}{D}\delta_{k,k'}\right|^2 \leq
  \frac{\|O\|_{\mathrm{op}}^2}{D}. \] holds for arbitrary bounded observables.
  If every term inside the sum has the same order, then the $D^{-1/2}$ scaling
  is naturally expected.}\BibitemShut {Stop}%
\end{thebibliography}

\end{document}